\documentclass[11pt]{article}  

\usepackage{graphicx}
\usepackage[utf8]{inputenc}
\usepackage[a4paper,left=2.5cm,right=2.5cm, bottom=2.5cm, top=2.5cm]{geometry}
\usepackage{amssymb}
\usepackage{amsmath}
\usepackage{bm}
\usepackage[margin=1pt,font=small,labelfont=bf]{caption}
\usepackage[dvipsnames]{xcolor}
\definecolor{citecolor}{RGB}{128,0,32}
\usepackage[unicode, pdfencoding=auto,psdextra,hyperfootnotes=false,breaklinks=true,colorlinks=true,allcolors=citecolor]{hyperref}
\usepackage{setspace}
\usepackage{fancyhdr}
\usepackage{titlesec}
\usepackage{algorithm}
\usepackage{algpseudocode}
\usepackage{etoolbox}

\usepackage{times}
\usepackage{tabularx}
\usepackage{datetime} 
\usepackage{lineno}
\usepackage{booktabs}
\usepackage{etoolbox}
\usepackage{subcaption}
\usepackage{longtable}  % in preamble

\patchcmd{\linenumber}{\hb@xt@}{\hbox}{}{}
\newdateformat{usvardate}{\shortmonthname[\THEMONTH]. \THEDAY, \THEYEAR}

\newcommand{\lastmodified}{%
  \textcolor{gray}{\scriptsize \usvardate\today{} at \currenttime}
}

\fancypagestyle{custom}
{      
\fancyhead[L]{}
\fancyhead[C]{\textcolor{gray}{\scriptsize revised manuscript \textit{Geophysical Research Letters}}}
\fancyhead[R]{}
\fancyfoot[C]{\textcolor{gray}{\thepage}}
\fancyfoot[R]{\lastmodified}
\fancyfoot[L]{}
}
\renewcommand\thesection{\arabic{section}}
\titlespacing\section{0pt}{12pt plus 4pt minus 2pt}{0pt plus 2pt minus 2pt}

\usepackage[
  giveninits=true,
  style=authoryear,
  backend=bibtex,
  maxcitenames=2,
  maxbibnames=6,
  hyperref=true,
  backref=false,
  sorting=nyt,
  uniquename=init,
  autolang=hyphen,
  dashed=false]{biblatex}

\AtEveryCite{\it \color{citecolor}}
\AtEveryBibitem{\clearfield{month}}
\DeclareGraphicsExtensions{.jpg}
\usepackage[breakable]{tcolorbox}
\definecolor{harsha}{RGB}{200,200,255}

\graphicspath{{./figs/}}
\newcommand{\sdrop}{1}

\newcommand{\brar}[1]{\left ( #1 \right )}
\newcommand{\bras}[1]{\left [ #1 \right ]}
\newcommand{\brac}[1]{\left \{ #1 \right \}}
\newcommand{\s}[1]{\sigma_{#1}}
\newcommand{\cut}[1]{\brar{z^{2}-a^{2}}^{#1}}
\newcommand{\phiz}{\Phi'(z)}
\newcommand{\phizz}{\Phi''(z)}
\newcommand{\prefac}{\dfrac{\sdrop}{2i}}
\newcommand{\ds}[1]{\Delta \sigma_{#1}}

\begin{document}

\pagestyle{custom}
\begin{center}
\LARGE {\bf Quantifying the Role of 3D Fault Geometry Complexities on Slow and Fast Earthquakes\\[12pt]}

\normalsize
J. Cheng$^{1*\P}$,
H. S. Bhat$^{1}$,
M. Almakari$^{1}$, 
B. Lecampion$^{2}$,
P. Dubernet$^{1}$

\begin{enumerate}
\small
\setlength\itemsep{-5pt}
\item Laboratoire de Géologie, Ecole Normale Superieure, CNRS-UMR 8538, 
PSL Research University, Paris, France\\
\item Geo-Energy Lab - Gaznat Chair on Geo-Energy, Swiss Federal Institute of Technology in Lausanne, \\EPFL-ENAC-IIC-GEL, Station 18, Lausanne CH-1015, Switzerland
\item[$*$] Currently at Division of Geological and Planetary Sciences, 
California Institute of Technology.
\item[$\P$] Corresponding author: \texttt{jcheng95@caltech.edu}
\end{enumerate}
\end{center}

\section*{Key Points}
\begin{itemize}
\item 3D simulations show interacting faults generate SSEs, earthquakes, or complex sequences, while a planar fault with identical frictional properties generates only earthquakes.
\item We quantify how strongly neighboring faults interact with each other through stress transfer (hereafter referred to as the fault interaction strength) and investigate how fault geometry controls the occurrence and proportion of slow slip events within slip sequences.
\item Observed SSE moment-duration scaling depends on the slip-rate detection threshold, suggesting instrumental sensitivity affects observations.
\end{itemize}

%%%%%%%%%%%%%%%%%%%%%%%%%%%%%%%%%%%%%%%%%%%%%%%%%%%%%%%%%%%%%%%%%%%%%%%%%%%%
%                           AUTHOR CONTRIBUTIONS
%%%%%%%%%%%%%%%%%%%%%%%%%%%%%%%%%%%%%%%%%%%%%%%%%%%%%%%%%%%%%%%%%%%%%%%%%%%%
\section*{CRediT}

\small

\begin{tabularx}{\textwidth}{rX}
\textbf{Conceptualization:} & H. S. Bhat \\
\textbf{Methodology:} & H. S. Bhat, J. Cheng \\
\textbf{Software:} & J. Cheng, B. Lecampion, P. Dubernet\\
\textbf{Investigation:} & J. Cheng, H. S. Bhat\\
\textbf{Writing - original draft:} & J. Cheng, H. S. Bhat\\
\textbf{Writing - review \& editing:} & H. S. Bhat, J. Cheng, B. Lecampion, M. Almakari\\
\textbf{Supervision:} &  H. S. Bhat, B. Lecampion, M. Almakari\\
\textbf{Funding acquisition:} &  H. S. Bhat
\end{tabularx}

%%%%%%%%%%%%%%%%%%%%%%%%%%%%%%%%%%%%%%%%%%%%%%%%%%%%%%%%%%%%%%%%%%%%%%%%%%%%
%                                ABSTRACT
%%%%%%%%%%%%%%%%%%%%%%%%%%%%%%%%%%%%%%%%%%%%%%%%%%%%%%%%%%%%%%%%%%%%%%%%%%%%
\section*{Abstract}
\small
Traditional models of slow slip events (SSEs) oversimplify fault geometry, although imaging shows subduction faults are segmented and complex. We examine how fault interactions control slip behavior using 3-D quasi-dynamic earthquake sequence simulations of two parallel faults with uniform rate-weakening friction accelerated by hierarchical matrices. Four regimes emerge—periodic earthquakes, coexisting SSEs and earthquakes, only SSEs, and complex sequences—whereas with the same friction condition a single planar fault produces only earthquakes. We quantify interaction using the maximum Coulomb stress induced on a target fault by a spatially uniform unit stress drop on a neighboring fault. Because the stress drop is normalized, the metric depends only on geometry and is independent of friction, allowing extension to arbitrary fault systems. SSEs occur only at intermediate fault interaction strengths. At low interaction strengths, the system produces regular, periodic earthquakes. At high interaction strengths, fault interactions generate complex earthquake sequences with irregular recurrence and variable magnitudes. Simulations reproduce observed moment–duration scaling and show sensitivity to detection thresholds. These results demonstrate geometric complexity alone generates both slow and fast earthquakes through evolving traction heterogeneity.

%%%%%%%%%%%%%%%%%%%%%%%%%%%%%%%%%%%%%%%%%%%%%%%%%%%%%%%%%%%%%%%%%%%%%%%%%%%%
%                                Plain Language Summary
%%%%%%%%%%%%%%%%%%%%%%%%%%%%%%%%%%%%%%%%%%%%%%%%%%%%%%%%%%%%%%%%%%%%%%%%%%%%
\section*{Plain Language Summary}
Slow slip events (SSEs) are slow fault movements that release stress gradually, with smaller stress drops and little to no seismic wave radiation, but they still change the stress on the fault and can influence where and when earthquakes occur. Traditional models often assume simple fault geometry, but real faults are complex and can interact with each other. Using 3D simulations of two parallel faults, we show that their basic interactions can generate a wide range of slip behaviors, including periodic earthquakes, slow slip events, and mixed or complex earthquake sequences. A single flat fault under the same conditions produces only earthquakes, highlighting the importance of fault geometry and interaction. We also find that the transition between slip behaviors depends on how strongly the faults interact each other. These results suggest that the complex geometry of real faults can naturally produce both slow and fast earthquakes observed in subduction zones.

\normalsize

%%%%%%%%%%%%%%%%%%%%%%%%%%%%%%%%%%%%%%%%%%%%%%%%%%%%%%%%%%%%%%%%%%%%%%%%%%%%
%                              INTRODUCTION
%%%%%%%%%%%%%%%%%%%%%%%%%%%%%%%%%%%%%%%%%%%%%%%%%%%%%%%%%%%%%%%%%%%%%%%%%%%%
\section{Introduction}

It has long been recognized that earthquake-related slip accounts for only a fraction of the overall slip budgets within plate tectonics. As continuous geodetic networks have improved, researchers have discovered slow slip events (SSEs) in various tectonic environments, for example in subduction zone: Cascadia subduction zone \parencite{hirose1999, dragert2001}, in continental strike-slip fault systems e.g., Haiyuan fault \parencite{Jolivet2013}, San Andreas fault \parencite{Rousset2019}. These events involve episodic, slow shear motion along faults (a few orders of magnitude faster than plate motion velocity) with no or minimal seismic activity. These events span a wide range of magnitudes, reaching up to approximately Mw 5–7.5, comparable to damaging earthquakes. Although SSEs generate little seismic radiation, large-magnitude events can still cause significant stress perturbations and accumulated slip on the fault, affecting its behavior. They are closely linked spatially and temporally with low-frequency earthquakes (LFEs), very low-frequency earthquakes (VLFEs), and tremors, exhibiting a lower frequency of seismic radiation compared to regular earthquakes of same magnitudes. Therefore, seismological instruments can indirectly detect slow slip events by tracking the migration of tremors \parencite{ito2007, shelly2007a, Michel2018}, repeating earthquakes \parencite{kato2012, Uchida2019}, or LFEs \parencite{bouchon2011,frank2019}, which improve detection capabilities for slow slip events.

Slow slip events are ubiquitous in subduction zones and exhibit a diverse range of spatiotemporal complexities. Sometimes, they can be observed in shallow depths or below the seismogenic zone. In Nankai Trough, short-term SSEs are discovered with a duration spanning from days to weeks and 3-6 months recurrence time \parencite{obara2004,hirose2006} and long-term SSEs are observed in deep areas with around a 1-year duration and 6-year recurrence time \parencite{Ozawa2001}. Shallow VLFEs and tremor, megathrust earthquake, long-term SSEs, and short-term SSEs are observed from trough to the deep regions of the subduction zone \parencite{obara2016}. This pattern of depth-dependent SSEs is also seen in the Mexican subduction zone \parencite{ElYousfi2023}. In Hikurangi, shallow SSEs are accompanied by microearthquakes, and deep SSEs are long-term with a duration of 2-3 months and a recurrence interval of 2 years with no tremors \parencite{wallace2013}. 

Slow slip events (SSEs) have a complex relationship with earthquakes in space and time. In the San Andreas fault, slow and fast rupture can coexist on the same section of the fault \parencite{Shelly2009,veedu2016}. Tremor signals were observed 18 months prior to the 2004 Mw 6.0 Parkfield earthquake. There are also examples of SSEs that can occur before earthquakes \parencite{burgmann2018,MartnezGarzn2024}. The 1999 Mw 7.6 Izmit earthquake was preceded by a 44-minute slow slip \parencite{bouchon2011}. A large slow slip event (equivalent Mw 7.6) in Guerrero triggered the 2014 Mw 7.3 Papanoa earthquake \parencite{radiguet2016}. In the Cascadia subduction zone, GPS observations suggest that the merging of slow slip event fronts (equivalent Mw 6.6) may be a possible mechanism for earthquake occurrence \parencite{Bletery2020}.

Earthquakes can also trigger SSEs. The 2016 Mw7.8 Kaikoura earthquake triggered a slow slip on the southern Hikurangi subduction zone \parencite{Wallace2018}. The 2017 Chiapas earthquake in Mexico triggered a slow slip event on the southern San Andreas Fault, located 3000 km away from the earthquake's epicenter \parencite{Tymofyeyeva2019}. SSEs can also occur periodically without earthquakes, like in Cascadia \parencite{rogers2003} and Hikurangi subduction zone \parencite{wallace2016}. The relationship between earthquakes and SSEs is still unclear and needs more studies and investigation. 

Those slow phenomena significantly influence fault behavior by altering the stress field and having intricate relationships with earthquakes \parencite{avouac2015,burgmann2018,obara2016}. Understanding slow slip events is crucial for gaining insights into earthquake mechanisms. There are several explanations for the mechanism of SSEs. SSEs can emerge from the transition of rate and state friction stability from velocity-weakening to velocity-strengthening \parencite{liu2005,rubin2008}. Heterogeneous frictional properties, such as varying the proportions of velocity-weakening and velocity-strengthening patches, can produce stable, slow, or dynamic slip events along the fault \parencite{skarbek2012,luo2017b,Nie2021}. Moreover, fault width plays a role in limiting rupture nucleation and stabilizing the faults \parencite{liu2007}. 2D simulations of two parallel faults with uniform rate-weakening friction shows geometry complexity can be a natural source of slow slip event \parencite{romanet2018}. Mechanisms like dilatant strengthening \parencite{segall2010a, liu2010} or frictional restrengthening at high slip speeds \parencite{kato2003, shibazaki2007, Im2021a} can modulate fault stabilization and instigate slow slip events. Additionally, thermal instabilities resulting from shear heating and temperature fluctuations can trigger SSEs \parencite{Wang2020}. Furthermore, the brittle-ductile (frictional and viscous deformation) transition \parencite{Nakata2011, Ando2023}, viscoelastic materials \parencite{Weng2025}, off-fault plastic deformation \parencite{Abdelmeguid2024, zhai2025, Mia2023}, and the presence of fluids \parencite{Bernaudin2018, cruz2018, Bhattacharya2019, Gao2017} are factors that also contribute to the occurrence of slow slip events. 

For a long time, subduction interfaces were assumed to be geometrically simple, particularly in studies of slow slip events focusing on heterogeneous friction or fluid effects. In reality, fault systems exhibit complex three-dimensional structures. For example, imaging of the Ecuadorian subduction zone \parencite{Chalumeau2024} shows earthquakes occurring across multi-fault segments and subparallel planes, challenging this view. This complexity underscores the need to incorporate realistic fault geometries when studying SSEs. Observations from Hikurangi, including drilling and seismic reflection data, further highlight the role of material and geometric complexity in promoting slow slip \parencite{Barnes2020,Kirkpatrick2021}. In Cascadia, downdip variability of SSEs suggests strong geometric control \parencite{Hall2018,mitsui2006}. Advanced simulations for Cascadia \parencite{liliu2016} and Guerrero \parencite{PerezSilva2021} indicate that non-planar geometries are critical for understanding SSEs. Geometric complexity is also central in models of precursory slow slip influenced by fault roughness \parencite{Cattania2021,Sun2025}. \textcite{romanet2018} first showed that interactions between neighboring faults influence earthquake cycle behavior and has been subsequently shown by other works \parencite[for e.g.][]{RodriguezPiceda2025, RodriguezPadilla2026}. Recent work \parencite{almakari2026} emphasizes the role of fault zone architecture in generating the full slip spectrum, and laboratory experiments \parencite{Kwiatek2024} show both fast and slow ruptures on rough faults. However, the role of stress interaction among multiple faults in 3D models of slow slip remains unresolved.

In this study, we investigate a step-over configuration of two parallel faults under spatially uniform, rate-weakening frictional conditions to explore the interplay between geometrical and frictional parameters in three dimensions. By considering homogeneous friction, we isolate the role of fault geometry as one possible mechanism for generating complex slip behavior. Three-dimensional simulations of fault slip are computationally intensive. Modeling slow slip events (SSEs) is particularly challenging because weakly rate-weakening friction produces large nucleation sizes, requiring more fault elements to maintain the same fault-length-to-nucleation-size ratio. In addition, fault interactions lead to complex slip-rate evolution, increasing the number of time steps needed to accurately resolve fault-slip behavior. To make such simulations feasible, we use a 3D quasi-dynamic earthquake sequence model based on the boundary element method accelerated with hierarchical matrices \parencite{Cheng2025,Lecampion_BigWham_a_C_2025}  . Our analysis identifies four distinct slip regimes: (1) only SSEs, (2) SSE-dominant behavior, (3) earthquake-dominant behavior, and (4) only earthquakes. Under identical parameters, a single planar fault produces only earthquakes (Figure B1), demonstrating that geometric complexity alone is sufficient to generate a broad range of slip behaviors, including SSEs.

We introduce a fault interaction metric, inspired by fracture mechanics, based on stress perturbations, fault width, length, overlap distance, and spacing, and use it to link geometric complexity with slip behavior through the ratio of SSE to total moment release. Moreover, our simulations reproduce the observed moment-duration scaling across different step-over geometries. The complex spatiotemporal slip patterns arise naturally from evolving traction heterogeneities, or “traction asperities” generated by fault interactions.

\section{Method}

We used the recently developed numerical approach, FASTDASH which integrates a 3D quasi-dynamic earthquake cycle modelling using boundary element methods accelerated by hierarchical matrices \parencite{Cheng2025}. This technique significantly optimizes computational efficiency, reducing complexity from $O(N^2)$ to $O(N \log{N})$, where $N$ represents the number of discretized fault elements. Such efficiency is crucial for solving complex 3D fault systems effectively. We analyse a fault system with two overlapping faults subjected to a far-field constant stress rate loading. Both faults are governed by laboratory-derived rate and state friction (RSF) law with aging state evolution. Friction is spatial uniform rate weakening. This model includes radiation damping to approximate inertial effect and neglects any wave propagation effect \parencite{rice1993}. Both shear and normal traction can vary with slip due to elastic interaction between two faults. With this approach, we can calculate key information on faults including maximum slip rate, moment rate, duration, and the distributions of stress and slip during multiple earthquake cycles. 

Linear stability analysis of rate and state friction brings out two important length scales namely the process zone size, $L_{b}$, and the nucleation length, $L_{nuc}$. $L_{b}$ represents the region where the strength breakdown occurs, and where traction and slip change rapidly. Numerical methods require a sufficient number of grid points within this zone to accurately capture these variations. In this study, we use a grid size of $\Delta s = L_{b}/3$. On the other hand, $L_{nuc}$ is the critical length necessary for slip instability to occur under idealized conditions. For 2D faults that governed by rate and state friction with ageing law, these parameters are defined as follows:
\begin{align}
L_{b} &= \frac{\mu D_{c}}{b\sigma_{n}}\\
L_{\text{nuc}} &= 
\begin{cases}
2.7548\,L_b, & 0 \le \dfrac{a}{b} \le 0.3781,\\[4pt]
\dfrac{2L_b}{\pi\,(1-a/b)^2}, & \dfrac{a}{b} \to 1~.
\end{cases}
% L_{nuc} &= 2 \times 1.3774 L_b ~~~0 \leq a/b \leq0.3781\\
% L_{nuc} &= \frac{2L_b}{\pi(1-a/b)^2} ~~~a/b \rightarrow 1
\end{align}
where $\mu$ is shear modulus, $D_{c}$ is the rate and state characteristic slip distance, a, b are the rate and state friction parameters to represent direct effect and evolution effect. For rate weakening friction, $0<a/b<1$. $\sigma_{n}$ is the normal traction \parencite{lapusta2009a, rubin2005, viesca2016b, Viesca2026}. The 3D nucleation length is larger than the 2D nucleation length by a factor of $\pi^2/4$ \parencite{chen2009, Viesca2026}.

We investigate the influence of geometric and frictional parameters on fault behavior in a three-dimensional compressional step-over configuration. The geometry is characterized by the fault width $W$, fault length $L_f$, overlap distance $L$, and separation distance $D$ between the two faults (Figure~1a). All length scales are normalized by the nucleation length $L_{\mathrm{nuc}}$, which depends on the frictional parameter ratio $a/b$. The fault aspect ratio is fixed at $L_f/W = 3$. We vary the normalized width $W/L_{\mathrm{nuc}}$ from 1.5 to 8 (corresponding to $L_f/L_{\mathrm{nuc}} = 4.5$--24), the overlap fraction $L/L_f$ from 0 to 1, and the normalized separation $D/L_{\mathrm{nuc}}$ from 0.1 to 5. Simulations are performed for $a/b \in \{0.4, 0.6, 0.8\}$.

We perform 65 simulations with different geometry and friction setting for our analysis. Tables in Appendix list the model parameters for all simulations.

In each simulation, an initial localized slip-rate perturbation is applied to nucleate the first earthquake, after which all subsequent events are included in the analysis. The simulations are run for sufficiently long durations to eliminate the influence of the initial condition. For consistent comparison across different frictional parameters, the total simulation time is chosen to correspond to the duration required for 10 earthquakes to occur on a planar fault with the same frictional properties.

\section{Results}

For a single fault that is longer than $L_{nuc}$ under spatially uniform rate-weakening friction ($a/b<1$), only fast earthquakes will occur \parencite{liu2005,rubin2008}. By considering a fault system with two parallel faults, spatio-temporal complex slip events emerge due to the time dependent evolution of traction heterogeneities produced by fault interaction (Figure 1).

We classify earthquakes and slow slip events based on their slip rates. An event is considered an earthquake if its maximum slip rate is greater than \(10^{-3} \, \text{m/s}\). The rupture on the secondary fault is triggered either by coseismic rupture jump or delayed renucleation. Slip rate of slow slip events is typically 1 to 2 order higher than plate rate. In our study, if the maximum slip rate falls between \(10^{-8} \, \text{m/s}\) and \(10^{-3} \, \text{m/s}\), it is identified as a slow slip event. Various slow slip events, as well as earthquakes having partial or full ruptures are identified in the synthetic catalog (Figure 1d). With various combinations of geometry and friction parameters, we identify four regimes: only slow slip events, SSE-dominant, earthquake-dominant and only earthquakes (Figure B2 and B3). We present the spatial-temporal complex events generated from our simulations and reproduce the moment-duration scaling law in this section. 

\subsection{Spatial-temporally complex events}
\begin{figure}
    \centering
\includegraphics[width=0.75\textwidth]{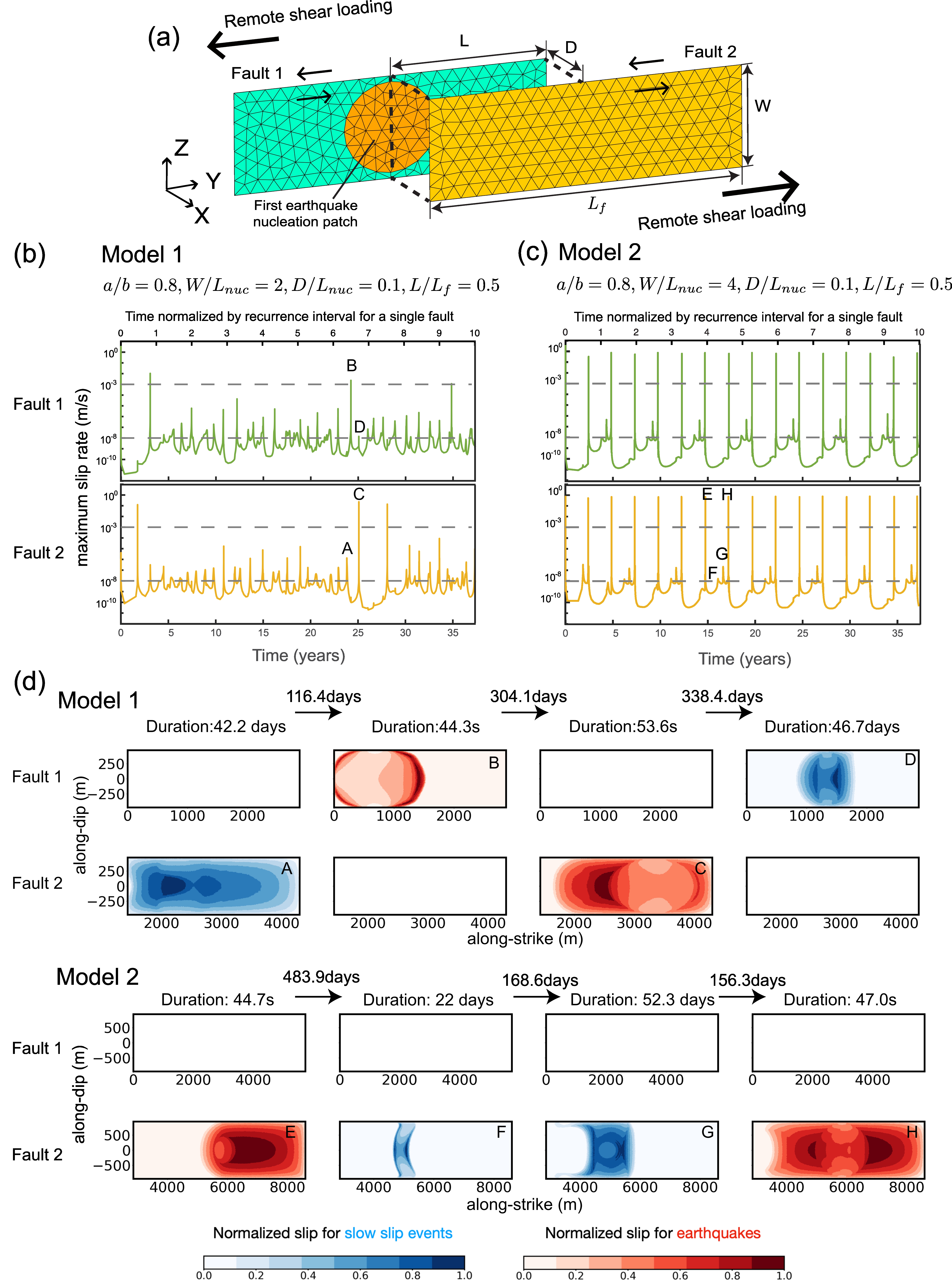}
\caption{(a) Step-over fault configuration. The mesh is shown exaggerated for visualization purposes; all simulations use a grid spacing of $L_b/3$ (b-c) Example of SSE-dominant regime and earthquake-dominant regime. Time evolution of the maximum slip rate. The bottom x-axis shows real time, while the top x-axis shows time normalized by the recurrence interval of a single fault with the same frictional properties. Green curve is on fault 1 and yellow curve is on fault 2. Grey dash lines are slip rate threshold for earthquakes and slow slip events (d) Spatial distribution of nomalized slip for slip sequences with selected events A, B, C, D in simulation showing in (b) and E, F, G, H in simulation showing in (c). Blue represents slow slip events and red represents earthquakes. Event A is slow slip event with full rupture. Event D, F and G are slow slip events with partial rupture. Event B and E are earthquakes with partial rupture. Event C and H are earthquakes with full rupture. Red and blue color shows normalized slip for earthquakes and slow slip events.}
\end{figure}

We show that the step-over configuration generates a wide range of spatiotemporally complex events, with both event durations and inter-event times varying substantially between events. To illustrate the transition between SSE-dominant and earthquake-dominant regimes, we discuss two examples: Model 1 and Model 2. The parameters are detailed in Table S1. Increasing the fault width from $2L_{nuc}$ to $4L_{nuc}$, together with the associated increases in fault length and overlap, shifts the system from SSE-dominant to earthquake-dominant behavior. In Model 1, slow slip events dominate the slip catalog, with only three earthquakes on each fault during the time span in which ten earthquakes occur on a single fault. In Model 2, we observe periodic earthquakes with two slow events occurring during the interseismic period. We selected four events from these two simulations to display the slip distribution on two fault planes (Fig. 1d), illustrating both SSEs and earthquakes occurring on the same faults with partial and full ruptures. Especially in Model 2, events E, F, G, and H form a chain of events, repeating on both faults. 

In the SSE-dominant regime, we observe full-rupture SSEs (Event A). In contrast, in the earthquake-dominant regime, SSEs are limited in extent (Events F and G), resulting in less moment released by SSEs in total catalog.  We also observe coalescence of two slow slip fronts and merge to a more energetic event (Figure B4).

By increasing fault width, the increased interaction between the faults results in larger stress perturbations, which in turn lead to more frequent earthquakes dominating the sequence of slip events. However, the interaction strength is also controlled by other geometrical parameters $L, D$ and $L_{f}$. We discuss the effect of other geometrical parameters in Section 4.1. 

\subsection{Moment-duration scaling law}

We perform seismic cycle simulations with various geometric and friction parameters, generating an extensive dataset in which slip events span several orders of magnitude in rupture area, slip, moment, and duration. 

In this study, events are identified using thresholds on the maximum slip rate of the fault system. We use a threshold of $10^{-3} m/s$ for earthquakes and $10^{-8} m/s$. An event starts when the maximum slip rate exceeds the corresponding threshold and ends when it drops below it. The event duration is defined as the time between these two points.

The moment is calculated from the accumulated slip $\delta$ during the duration over the fault area during the event. ${M} = \int_{A}\mu \delta dA$, where $\mu$ is the shear modulus and $A$ is the fault area. To avoid numerical noise, we discard events that last fewer than five time steps.

We analyzed slip sequences from each simulation over an equivalent time period (the duration required for a single fault to have 10 earthquakes). We generated moment-duration scaling plots for both slow slip events (SSEs) and earthquakes (EQs), revealing different scaling behaviors: earthquakes exhibit cubic scaling, whereas SSEs are bounded by linear scaling.

Our results show that the moment--duration scaling of slow slip events depends strongly on how the events are defined. Lower slip-rate thresholds detect more events and include slower portions of slip, which increases the inferred duration and moment and modifies the scaling relation. Using a threshold of \(10^{-8} m/s\), we identify 1925 SSEs with a scaling \(M \sim T^{1.2}\). Increasing the threshold to \(10^{-6} m/s\) reduces the catalog to 640 events and steepens the scaling to \(M \sim T^{1.5}\). Because the scaling is fitted using the entire population of slow slip events, the resulting exponent represents the average scaling behavior of all slow slip events.

Looking in more detail, large SSEs (moment \(10^{12}\text{--}10^{16} N\cdot m\)) approach cubic scaling. A possible explanation is that small, slow events are dominated by slip accumulation over a nearly fixed area, leading to near-linear scaling, whereas larger events involve significant rupture growth and therefore approach cubic scaling due to three-dimensional effects. The wide scatter likely reflects event-to-event variability in rupture velocity.

These results highlight the sensitivity of SSE detection and inferred scaling to the chosen slip-rate threshold and emphasize the need for consistent event definitions when comparing SSE dynamics across studies \parencite{almakari2026,Costantino2026}.

\begin{figure}
    \centering
\includegraphics[width=0.75\textwidth]{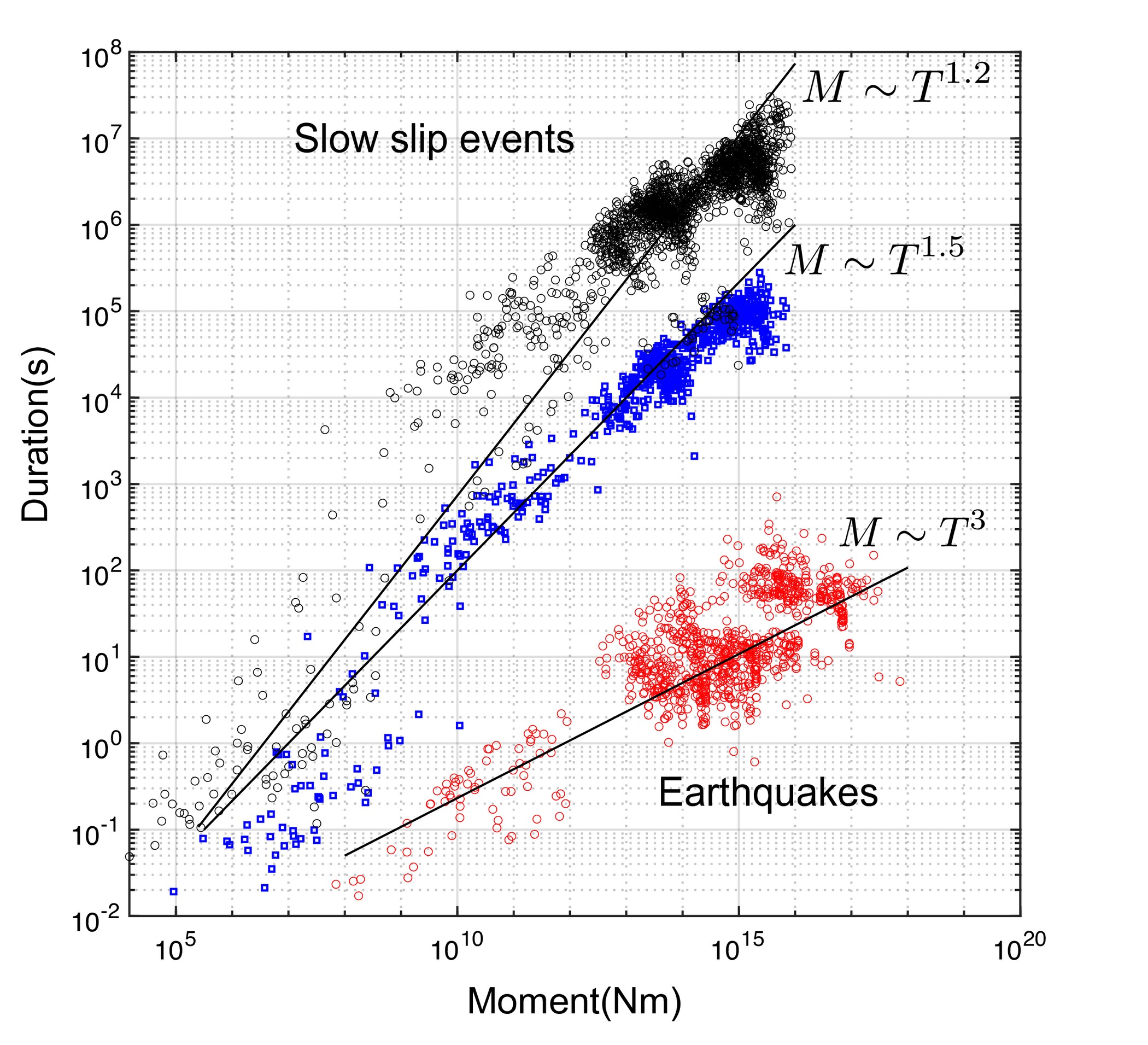}
\caption{Moment-Duration scaling across all simulations. Red denotes earthquakes exhibiting cubic scaling. Blue and black represent slow slip events identified with slip rate thresholds of \(10^{-6} m/s\) and \(10^{-8} m/s\), respectively, showing varying but predominantly linear scaling from \(M \sim T^{1.5}\) to \(M \sim T^{1.2}\)}
\end{figure}

For earthquakes, the scaling is insensitive to the detection threshold and consistently follows cubic scaling. The observed scatter may result from variations in rupture velocity between events.

\section{Discussion}

\subsection{A metric to quantify fault interaction}

The coexistence of slow slip events (SSEs) and earthquakes (EQs) is controlled by stress perturbations from neighboring faults. However, no quantitative framework currently links geometrical complexity—overlap $L$, separation $D$, width $W$, and length $L_f$—to the emergence of slow and fast slip. 

We introduce a dimensionless interaction metric $\Lambda$ that quantifies the maximum coulomb stress change transferred from a primary to a secondary fault. The metric is derived from 2D fracture mechanics for a shear fracture subject to unit stress drop using Muskhelishvili–Kolosov potentials \parencite{scheel2022}, yielding
\[
\Lambda = f(L/L_f, D/L_f),
\]
with the full derivation in Appendix A. Unlike linear elastic fracture mechanics (where solutions hold for small distances from the crack) the complex potentials take the full field into account. Because crack interaction scales linearly with stress drop, $\Lambda$ directly measures interaction strength:
\begin{align}
\Lambda &= \textrm{max}\brac{-2\Im{\bras{\phiz}} - 2y\Re{\bras{\phizz}} + 2f_{s}y\Im{\bras{\phizz}}}
\end{align}
where $f_{s} = 0.6$ is assumed, $z \in [x_1+iD, x_2+iD]$ and \(x_1 = (2L - 1)L_f/2\), \(x_2 = x_1 + L_f\). Geometrical parameters are shown in Fig 3(a).

\begin{figure}[h!]
    \centering
\includegraphics[width=0.95\textwidth]{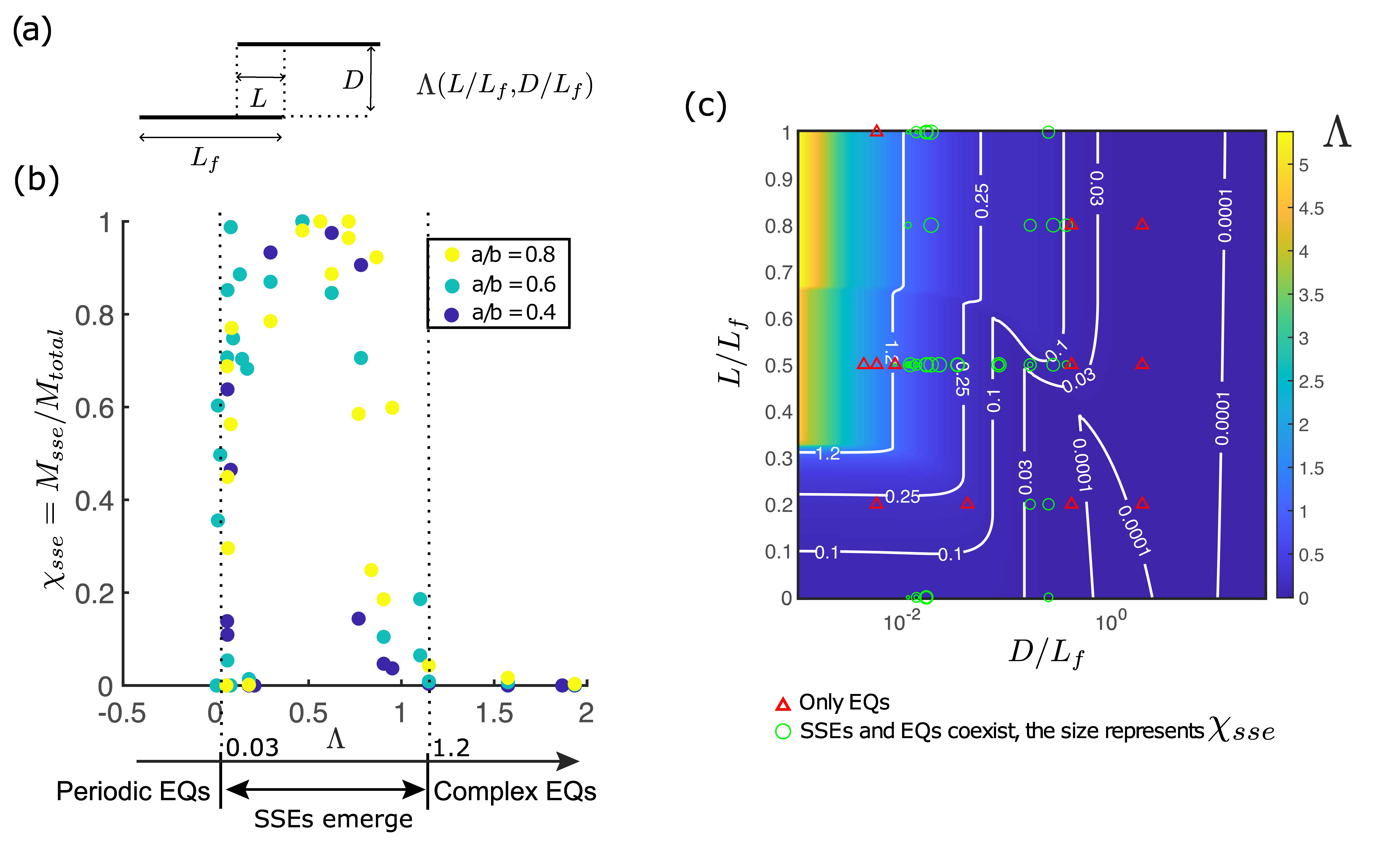}
\caption{(a) Definition of $\Lambda$ in terms of overlap $L$, separation $D$, and fault length $L_f$. 
(b) Relationship between the interaction metric $\Lambda$ and the SSE ratio $\chi_{sse}$. Colors denote different $a/b$. 
(c) $\Lambda$ as a function of normalized overlap and separation. Red triangles indicate EQ-only simulations; green circles indicate coexistence, with size proportional to $\chi_{sse}$.}
\end{figure}

Within our parameter space, $\Lambda$ ranges from 0 to 2.2 (Fig.~3b). Small values correspond to weak interaction and independent slip, whereas large values reflect strong stress coupling. We quantify slip style using the SSE ratio $\chi_{sse}$, defined as the fraction of moment released aseismically over ten earthquake cycles. $\chi_{sse}=0$ indicates purely seismic slip, $\chi_{sse}=1$ purely aseismic.

Geometry strongly controls slip partitioning (Fig.~3b). When $\Lambda<0.03$ or $\Lambda>1.2$, faults are earthquake-dominated. Low $\Lambda$ produces periodic earthquakes (Fig.~S2), whereas high $\Lambda$ yields complex ruptures (Fig.~S3). At low $\Lambda$, separation $D$ dominates, with $\Lambda \propto 1/\sqrt{D}$. At high $\Lambda$, width $W$ dominates, with $\Lambda \propto \sqrt{W}$.

In contrast, SSEs and EQs coexist for \(0.03<\Lambda<1.2\), with SSE-dominant behavior for \(0.05<\Lambda<0.75\). SSEs are highly sensitive to stress perturbations \parencite{obara2016}, requiring intermediate interaction strength: too weak and perturbations are insufficient; too strong and slow slip evolves into earthquakes.

Although varying $a/b$ slightly shifts regime boundaries (Fig.~3b), geometry remains the primary control. Different combinations of overlap and separation can produce identical $\Lambda$ (Fig.~3c). $\Lambda$ decreases rapidly with increasing separation and increases with overlap, though the overlap effect weakens at large distances. Most coexistence simulations fall within the intermediate $\Lambda$ range, and slip behavior transitions smoothly toward purely seismic regimes near the thresholds. This behavior is insensitive to friction (Fig.~3b), highlighting geometry as the primary control on slip complexity. 

We introduce $\Lambda$ as a physically motivated interaction metric that links fault geometry to the emergence of slow and fast slip events, showing that fault interactions are governed by stress transfer rather than geometric distance alone. Future work should extend this metric to more realistic fault systems.

\subsection{Traction asperities}

Notably, our study assumes spatially uniform rate-weakening friction. Previous single planar fault studies \parencite{Lay2012,veedu2016} showed that slow slip events emerge in small rate-weakening asperities, whereas earthquakes occur in larger ones. These frictional asperities are spatially and temporally stable. Here, we introduce the concept of traction asperities, which arise naturally in multi-fault systems from stress interactions and prior events. These asperities also control the nucleation and arrest of both slow and fast earthquakes.

Figure~4 shows the shear-to-normal traction ratio on Fault~2 before and after selected events (E–H) in Model~2 (Figure~1). Due to stress interactions with neighboring faults and previous ruptures, the traction field becomes spatially and temporally heterogeneous. For example, the left edge overlapping Fault~1 develops a low traction-ratio patch. Nucleation occurs in regions with a high shear-to-normal traction ratio (HRA). When the HRA is narrow, nucleation may not complete within a single asperity, leading to aseismic events (e.g., F and G). In contrast, regions with a low traction ratio (LRA) are stronger and act as barriers. After nucleation, event~E propagates toward the right HRA and arrests at the left LRA, producing a partial rupture; events F and G behave similarly. Heterogeneous traction complicates nucleation, whereas coalescing slip fronts and more homogeneous traction can produce full ruptures, as in event~H.

\begin{figure}[t!]
    \centering
\includegraphics[width=0.75\textwidth]{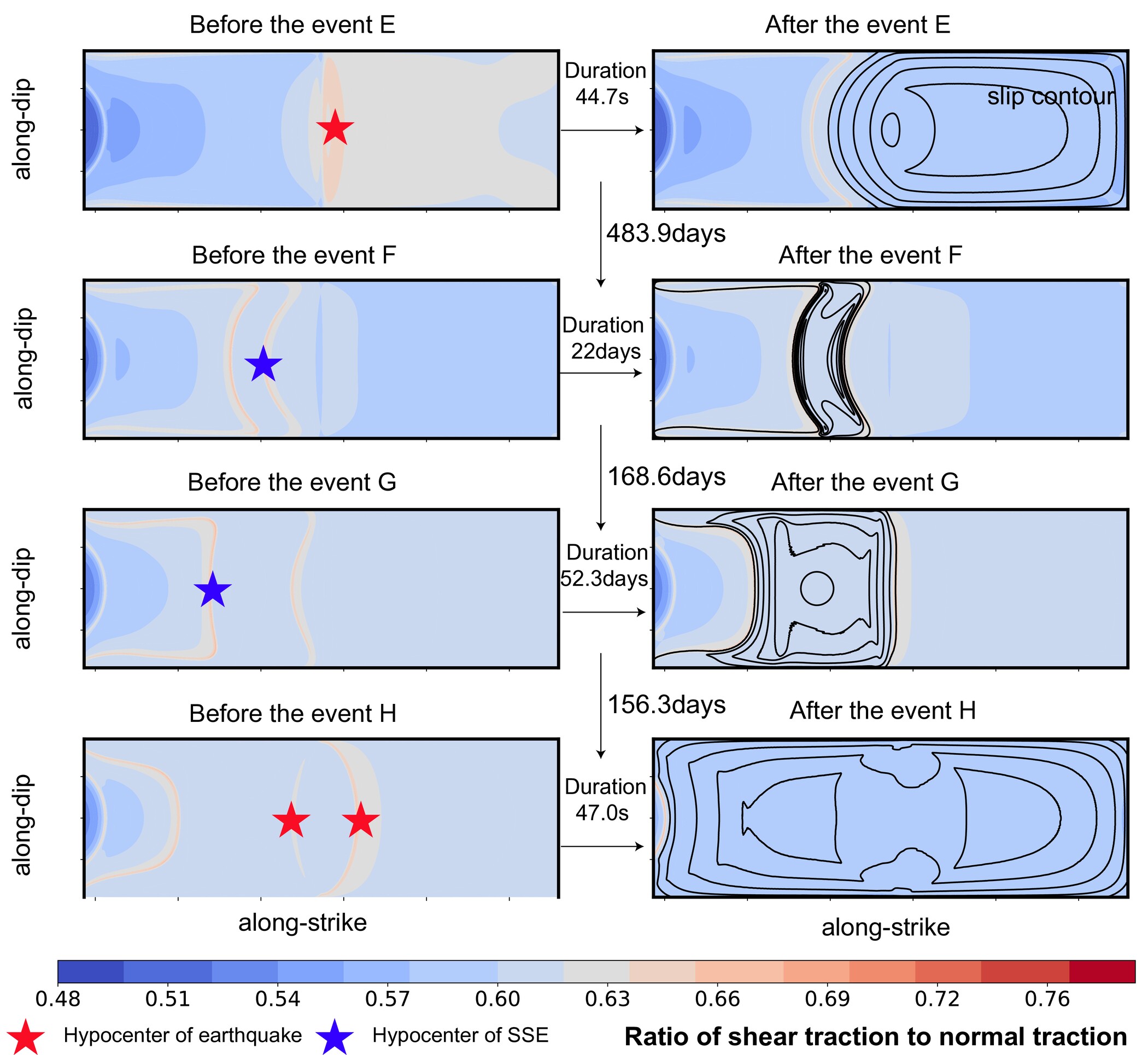}
\caption{Traction field evolution during four events (E, F, G, and H) on fault 2 in Model 2. The colormap indicates the ratio of shear to normal traction. The first column displays the traction field before the nucleation of each event, with red stars marking earthquake hypocenters and blue stars indicating SSE hypocenters. The second column shows the traction field after each event, with black contour lines representing the slip distribution. The duration of each event and the inter-event times are noted.}
\end{figure}

% \subsection{Compare with 2D models}

% Compared to the 2D method used in \textcite{romanet2018}, our 3D method does not rely on the assumption of a fixed rupture depth of 1 km when calculating the moment to reproduce scaling laws. Because the 2D model lacks the third dimension, they have to assume the length in that dimension, which is totally arbitrary. The 3D method allows for a more accurate understanding of scaling relationships. Our results show that moment-duration scaling is sensitive to the slip rate threshold for slow slip events, a sensitivity that previous 2D models did not reveal. 

\subsection{Comparison with single planar fault with heterogeneous friction properties}

We examine a system of two parallel faults and show that their stress interaction generates a spatio-temporal stress field that produces slow slip events and earthquakes with partial and full ruptures. This complexity emerges from fault interaction and rupture history, rather than fixed fault properties. 

In contrast, models based on frictional heterogeneity \parencite{dublanchet2013,Kaneko2010} attribute complex behavior to static variations in frictional parameters, which require tuning asperity size and distribution to reproduce observations. Our results demonstrate that even simple geometric configurations can naturally generate diverse moment–duration scaling and complex nucleation patterns consistent with natural slip. 

Because geometry can be constrained by surface mapping and seismic imaging, it provides a dynamically evolving and observationally accessible framework for interpreting complex slip. Integrating geometric and frictional effects will further improve earthquake models.

\section{Conclusion}

In this study, we investigate the emergence of complex slow and fast seismic events driven by elastic interactions and heterogeneous stress fields in a 3D fault system composed of two parallel planar faults. Using a 3D quasi-dynamic earthquake cycle model based on the boundary element method accelerated by hierarchical matrices \parencite{Cheng2025}, we explore how fault geometry and frictional properties control slip dynamics. For a single isolated fault under spatially uniform rate-weakening friction, only fast earthquakes occur when the fault length exceeds the nucleation length. However, when fault interactions are included, stress coupling between neighboring faults generates spatiotemporally complex slip behaviors, including slow slip events (SSEs) and partial or full earthquake ruptures. 

We quantify geometric complexity using a fault interaction metric defined as the maximum Coulomb stress change induced on a target fault by unit, spatially uniform stress drop on a neighboring fault. Slip complexity is characterized using the SSE moment release ratio, defined as the fraction of total seismic moment released during slow slip events. SSEs occur only within an intermediate range of interaction strength—too weak or too strong coupling promotes earthquake-dominated behavior—and this pattern remains robust across frictional parameters. Our results also reproduce the moment-duration scaling of SSEs, showing a predominantly linear trend that depends strongly on the slip rate threshold used to identify events. These findings highlight how 3D fault interactions can naturally generate the coexistence of earthquakes and SSEs, bridging the gap between seismic and aseismic slip behaviors.

%%%%%%%%%%%%%%%%%%%%%%%%%%%%%%%%%%%%%%%%%%%%%%%%%%%%%%%%%%%%%%%%%%%%%%%%%%%%
%                           ACKNOWLEDGEMENTS
%%%%%%%%%%%%%%%%%%%%%%%%%%%%%%%%%%%%%%%%%%%%%%%%%%%%%%%%%%%%%%%%%%%%%%%%%%%%
\section*{Acknowledgements}

JC, HSB and MA gratefully acknowledge the European Research Council (ERC) for its full support of
this work through the PERSISMO grant (No. 865411). The numerical simulations presented in this study
were performed on the MADARIAGA cluster, also supported by the ERC PERSISMO grant. 

%%%%%%%%%%%%%%%%%%%%%%%%%%%%%%%%%%%%%%%%%%%%%%%%%%%%%%%%%%%%%%%%%%%%%%%%%%%%
% CONFICT OF INTEREST
%%%%%%%%%%%%%%%%%%%%%%%%%%%%%%%%%%%%%%%%%%%%%%%%%%%%%%%%%%%%%%%%%%%%%%%%%%%%

\section*{Conflict of Interest}
The authors declare no conflicts of interest relevant to this study.
%%%%%%%%%%%%%%%%%%%%%%%%%%%%%%%%%%%%%%%%%%%%%%%%%%%%%%%%%%%%%%%%%%%%%%%%%%%%
%                           DATA AVAILABILITY
%%%%%%%%%%%%%%%%%%%%%%%%%%%%%%%%%%%%%%%%%%%%%%%%%%%%%%%%%%%%%%%%%%%%%%%%%%%%
\section*{Open Research Section}
The datasets generated and analyzed during this study, and codes used to analyze, are available at \textcite{cheng2026b}.

%%%%%%%%%%%%%%%%%%%%%%%%%%%%%%%%%%%%%%%%%%%%%%%%%%%%%%%%%%%%%%%%%%%%%%%%%%%%
%                      APPENDICES
%%%%%%%%%%%%%%%%%%%%%%%%%%%%%%%%%%%%%%%%%%%%%%%%%%%%%%%%%%%%%%%%%%%%%%%%%%%%

\setcounter{section}{0}
\renewcommand\thesection{Appendix \Alph{section}}
\renewcommand\thesubsection{\Alph{section}\arabic{subsection}}
\renewcommand{\theequation}{\Alph{section}\arabic{equation}}
\setcounter{equation}{0}
\renewcommand{\thefigure}{\Alph{section}\arabic{figure}}
\setcounter{figure}{0}

% Fix hyperref identifiers to avoid duplicates
\renewcommand{\theHsection}{appendix.\Alph{section}}
\renewcommand{\theHsubsection}{appendix.\Alph{section}.\arabic{subsection}}
\renewcommand{\theHequation}{appendix.\Alph{section}.\arabic{equation}}

%%%%%%%%%%%%%%%%%%%%%%%%%%%%%%%%%%%%%%%%%%%%%%%%%%%%%%%%%%%%%%%%%%%%%%%%%%%%
%                      APPENDIX A
%%%%%%%%%%%%%%%%%%%%%%%%%%%%%%%%%%%%%%%%%%%%%%%%%%%%%%%%%%%%%%%%%%%%%%%%%%%%
\section{The definition of fault interaction metric}

The Muskhelisvili-Kolosov complex potentials for a shear crack of length $2a$ are given by \parencite{scheel2022}
\begin{align}
\phiz &= \prefac\bras{\dfrac{z}{\cut{1/2}}-1}\\
\phizz &= \prefac\bras{\dfrac{-a^{2}}{\cut{3/2}}}\\
\Psi'(z) &= -2\phiz-z\phizz
\end{align}

The stress field is then given by $\s{ij}(z) = \s{ij}^{0} + \ds{ij}(z)$ where
The stress field is then given by $\s{ij}(z) = \s{ij}^{0} + \ds{ij}(z)$ where
%\begin{align}
%\ds{11}(z)+\ds{22}(z) &= 4\Re{\bras{\phiz}} \\
%\ds{22}(z)-\ds{11}(z)+2i\ds{12}(z) &= 2\bras{\overline{z}\phizz + \Psi'(z)} = -4\bras{\phiz+iy\phizz}
%\end{align}
%
% Thus,
\begin{align}
\ds{22}(z) &= 2y\Im{\bras{\phizz}}\\
\ds{12}(z) &= -2\Im{\bras{\phiz}} - 2y\Re{\bras{\phizz}}
\end{align}
where $\Re$ and $\Im$ correspond to the real and imaginary parts of their arguments respectively. See Figure B5. Length $2a$ is the minimum value of fault length $L_{f}$ and fault width $W$. The complex coordinate of fault 2 is denoted as $z$.

Fault 1 spans from \((-L_f/2, 0)\) to \((L_f/2, 0)\). Fault 2, on the other hand, extends from \((x_1, D)\) to \((x_2, D)\), where \(x_1 = (2L - 1) \cdot (L_f/2)\), \(x_2 = x_1 + L_f\) and $L$ is the overlap.

We define the metric $\Lambda$ as the maximum of \(\Delta \sigma_{12} + f_{s}\Delta \sigma_{22}\) on fault 2 (Figure B6), representing the maximum Coulomb stress transfer from one fault to the other. This metric, which is a function of all geometrical parameters, quantifies the strength of the stress interaction between the two faults.  
\begin{align}
\Lambda &= \textrm{max}\brac{-2\Im{\bras{\phiz}} - 2y\Re{\bras{\phizz}} + 2f_{s}y\Im{\bras{\phizz}}}
\end{align}
where $f_{s} = 0.6$ is assumed.

\section{Method: FASTDASH}

The simulations in this study are performed using FASTDASH (FAult SysTem Dynamics: Accelerated Solver using H-matrices), a three-dimensional quasi-dynamic earthquake cycle simulator based on the boundary element method (BEM). A complete description and validation of the method are provided in \textcite{Cheng2025}; here we summarize the main features relevant to this study.

The fault surfaces are discretized into triangular boundary elements. Fault slip is governed by rate-and-state friction with the aging law. The quasi-dynamic approximation accounts for radiation damping while neglecting inertial wave propagation, allowing efficient simulation of long-term earthquake cycles.

Elastic interactions between all fault elements are computed using the BEM. To reduce the computational cost associated with the dense influence matrix, FASTDASH employs hierarchical matrices (H-matrices), which compress the elastic kernels and reduce the computational complexity from $O(N^2)$ to approximately $O(NlogN)$. This enables simulations of large three-dimensional fault systems over many earthquake cycles.

The governing ordinary differential equations are integrated using an adaptive Runge--Kutta 45 time-stepping scheme, which automatically adjusts the time step according to the evolving slip rate. This approach efficiently resolves both long interseismic periods and rapid coseismic slip within a unified framework.

For the simulations presented in this study, all models use spatially homogeneous rate-and-state friction, and only the fault geometry is varied to investigate the role of elastic fault interactions in generating slow and fast slip events.

\begin{table}[h!]
\centering
\caption{Model and geometric parameters used in the simulations.}
\begin{tabular}{l l c c}
\hline
\multicolumn{4}{c}{\textbf{Friction and material parameters}}\\
\hline
Parameter & Description & \multicolumn{2}{c}{Value} \\
\hline
$a$ & RSF direct effect parameter & \multicolumn{2}{c}{0.008} \\
$b$ & RSF evolution effect parameter & \multicolumn{2}{c}{0.01} \\
$D_{c}$ & Characteristic slip distance & \multicolumn{2}{c}{0.001 m} \\
$V_{ref}$ & Reference slip rate & \multicolumn{2}{c}{$10^{-6}$ m/s} \\
$f_{0}$ & Reference friction coefficient & \multicolumn{2}{c}{0.6} \\
$\mu$ & Shear modulus & \multicolumn{2}{c}{30 GPa} \\
$\rho$ & Density & \multicolumn{2}{c}{2670 kg m$^{-3}$} \\
$C_{s}$ & Shear wave velocity & \multicolumn{2}{c}{3464 m/s} \\
$\nu$ & Poisson's ratio & \multicolumn{2}{c}{0.25} \\
$V_{0}$ & Initial slip rate & \multicolumn{2}{c}{$10^{-9}$ m/s} \\
$\dot{\tau}_{s}$ & Shear loading rate & \multicolumn{2}{c}{0.05 Pa/s} \\

\hline
\multicolumn{4}{c}{\textbf{Geometric parameters}}\\
\hline
Parameter & Description & Model 1 & Model 2 \\
\hline
$L_{nuc}$ & Nucleation length & 477.46 m & 477.46 m \\
$D$ & Fault separation & 47.746 m & 47.746 m \\
$W$ & Fault width & 954.92 m & 1909.80 m \\
$L_{f}$ & Fault length & 2864.80 m & 5729.50 m \\
$L$ & Overlap distance & 1432.40 m & 2864.80 m \\
\hline
\end{tabular}
\label{tab:parameters}
\end{table}

\clearpage

\begin{longtable}{ccccc}
\hline
ID & $a/b$ & $W/L_\mathrm{nuc}$ & $D/L_\mathrm{nuc}$ & $L/L_f$ \\
\hline
\endfirsthead

\multicolumn{5}{c}{\tablename~\thetable{} -- \textit{continued from previous page}} \\
\hline
ID & $a/b$ & $W/L_\mathrm{nuc}$ & $D/L_\mathrm{nuc}$ & $L/L_f$ \\
\hline
\endhead

\hline
\multicolumn{5}{r}{\textit{continued on next page}} \\
\endfoot

\hline
\endlastfoot

1  & 0.4 & 1.5  & 0.1  & 0.5 \\
2  & 0.4 & 2.5  & 0.1  & 0.0 \\
3  & 0.4 & 2.5  & 0.1  & 0.5 \\
4  & 0.4 & 2.5  & 0.1  & 1.0 \\
5  & 0.4 & 2.0  & 0.1  & 0.0 \\
6  & 0.4 & 2.0  & 0.1  & 0.5 \\
7  & 0.4 & 2.0  & 0.1  & 1.0 \\
8  & 0.4 & 2.0  & 0.2  & 0.5 \\
9  & 0.4 & 2.0  & 0.5  & 0.5 \\
10 & 0.4 & 2.0  & 1.0  & 0.5 \\
11 & 0.4 & 3.0  & 0.1  & 0.5 \\
12 & 0.4 & 4.0  & 0.1  & 0.5 \\
13 & 0.4 & 4.0  & 0.5  & 0.2 \\
14 & 0.4 & 4.0  & 5.0  & 0.5 \\
15 & 0.4 & 6.0  & 0.1  & 0.5 \\
16 & 0.4 & 6.0  & 0.1  & 0.2 \\
17 & 0.4 & 6.0  & 0.1  & 1.0 \\
18 & 0.4 & 8.0  & 0.1  & 0.5 \\
19 & 0.6 & 1.5  & 0.1  & 0.5 \\
20 & 0.6 & 1.8  & 1.5  & 0.8 \\
21 & 0.6 & 1.8  & 2.0  & 0.8 \\
22 & 0.6 & 2.0  & 0.1  & 0.0 \\
23 & 0.6 & 2.0  & 0.1  & 0.5 \\
24 & 0.6 & 2.0  & 0.1  & 1.0 \\
25 & 0.6 & 2.0  & 0.2  & 0.5 \\
26 & 0.6 & 2.0  & 0.5  & 0.5 \\
27 & 0.6 & 2.0  & 1.5  & 0.0 \\
28 & 0.6 & 2.0  & 1.5  & 0.2 \\
29 & 0.6 & 2.0  & 1.5  & 1.0 \\
30 & 0.6 & 2.0  & 1.0  & 0.2 \\
31 & 0.6 & 2.0  & 1.0  & 0.5 \\
32 & 0.6 & 2.0  & 1.0  & 0.8 \\
33 & 0.6 & 3.0  & 0.1  & 0.0 \\
34 & 0.6 & 3.0  & 0.1  & 0.5 \\
35 & 0.6 & 3.0  & 0.1  & 0.8 \\
36 & 0.6 & 3.0  & 0.1  & 1.0 \\
37 & 0.6 & 4.0  & 0.1  & 0.5 \\
38 & 0.6 & 4.0  & 0.5  & 0.2 \\
39 & 0.6 & 4.0  & 5.0  & 0.5 \\
40 & 0.6 & 4.0  & 5.0  & 0.8 \\
41 & 0.6 & 4.0  & 5.0  & 0.2 \\
42 & 0.6 & 6.0  & 0.1  & 0.5 \\
43 & 0.6 & 8.0  & 0.1  & 0.5 \\
44 & 0.8 & 1.5  & 0.1  & 0.5 \\
45 & 0.8 & 1.8  & 0.1  & 0.5 \\
46 & 0.8 & 1.8  & 0.1  & 0.8 \\
47 & 0.8 & 1.8  & 0.1  & 1.0 \\
48 & 0.8 & 1.8  & 1.5  & 0.5 \\
49 & 0.8 & 1.8  & 2.0  & 0.5 \\
50 & 0.8 & 2.5  & 0.1  & 0.0 \\
51 & 0.8 & 2.5  & 0.1  & 0.5 \\
52 & 0.8 & 2.5  & 0.1  & 1.0 \\
53 & 0.8 & 2.75 & 0.1  & 0.5 \\
54 & 0.8 & 2.0  & 5.0  & 0.5 \\
55 & 0.8 & 2.0  & 0.07 & 0.5 \\
56 & 0.8 & 2.0  & 0.1  & 0.5 \\
57 & 0.8 & 2.0  & 0.2  & 0.5 \\
58 & 0.8 & 2.0  & 0.5  & 0.5 \\
59 & 0.8 & 2.0  & 1.0  & 0.5 \\
60 & 0.8 & 3.0  & 0.1  & 0.5 \\
61 & 0.8 & 4.0  & 0.1  & 0.5 \\
62 & 0.8 & 4.0  & 0.5  & 0.2 \\
63 & 0.8 & 4.0  & 5.0  & 0.5 \\
64 & 0.8 & 6.0  & 0.1  & 0.5 \\
65 & 0.8 & 8.0  & 0.1  & 0.5 \\

\caption{Simulation parameters.}
\label{tab:simparams}
\end{longtable}

\begin{figure}[h!]
    \centering
    \includegraphics[width=0.65\textwidth]{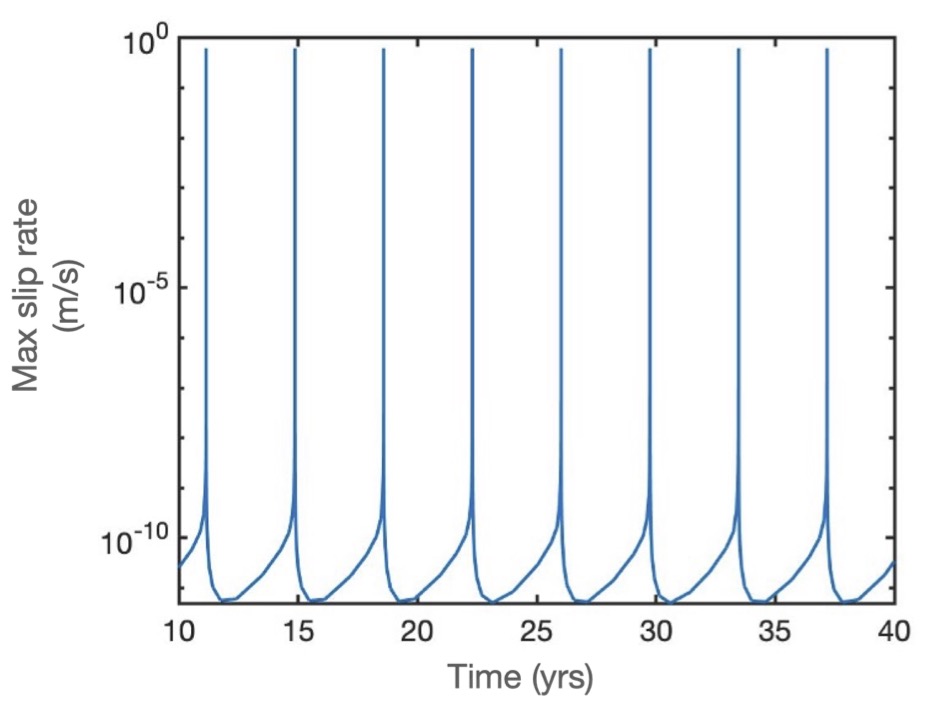}
    \caption{A single-fault model exhibits periodic earthquakes ($a/b = 0.8$,
    $W/L_\mathrm{nuc} = 2$) with a recurrence interval of 3.73 years.}
    \label{fig:S1}
\end{figure}

\begin{figure}[h!]
    \centering
    \includegraphics[width=0.65\textwidth]{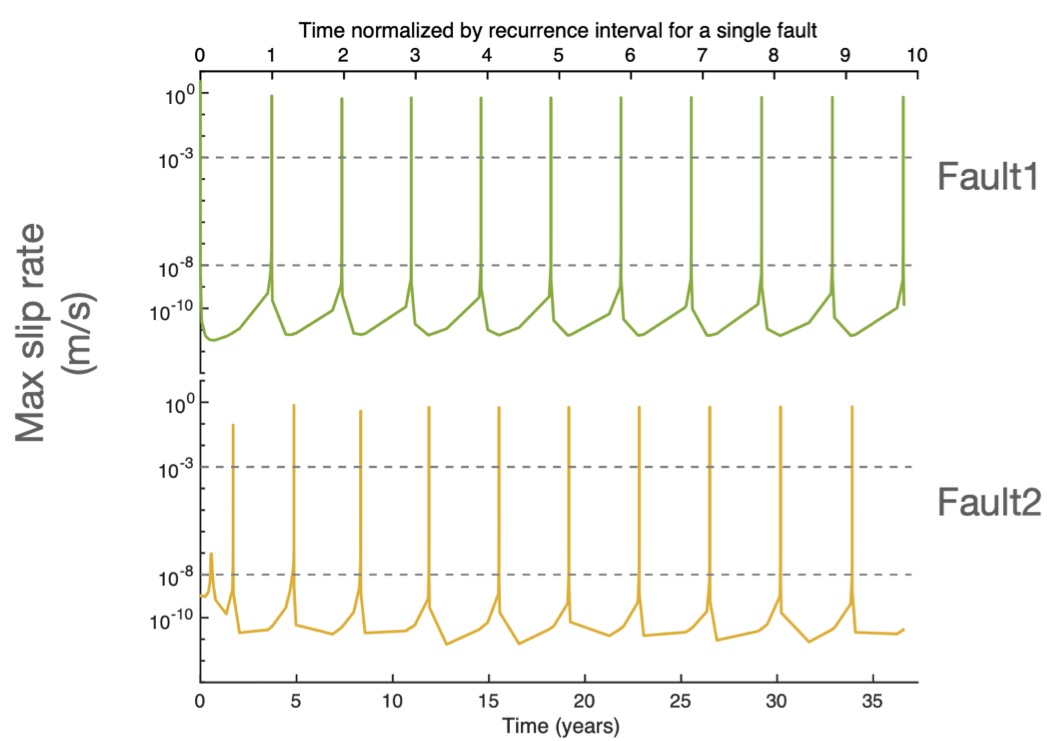}
    \caption{Periodic earthquake regime with $\lambda = 0.0183$ for the simulation
    with $a/b = 0.8$, $W/L_\mathrm{nuc} = 2$, $D/L_\mathrm{nuc} = 5$, and
    $L/L_f = 0.5$.}
    \label{fig:S2}
\end{figure}

\begin{figure}[htbp]
    \centering
    \includegraphics[width=0.75\textwidth]{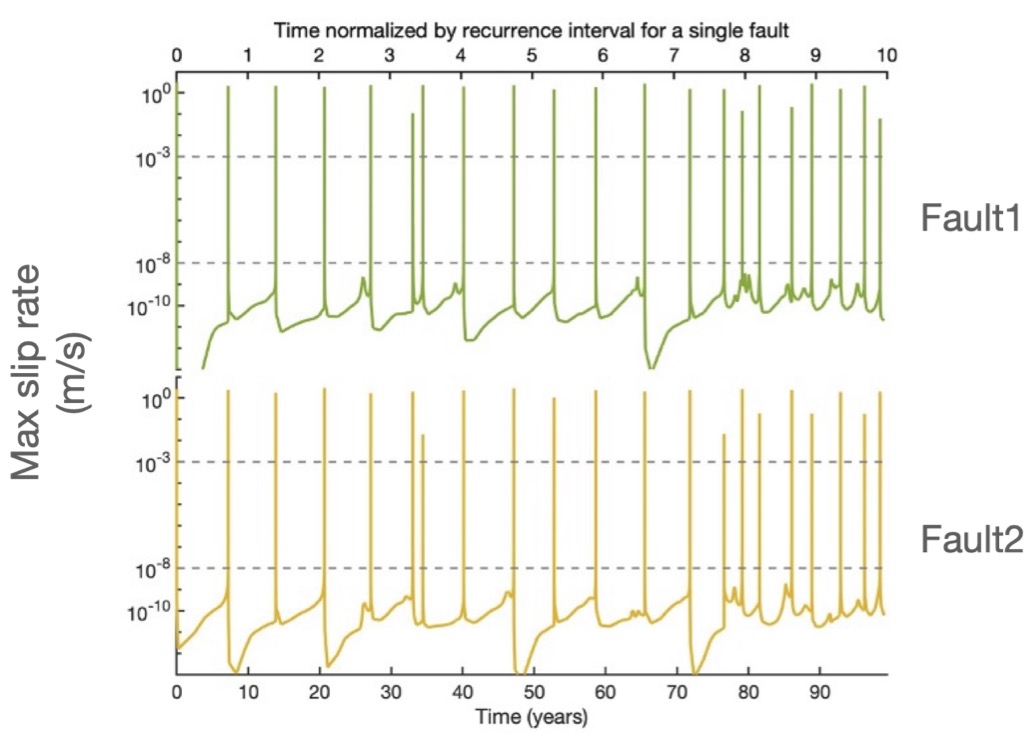}
    \caption{Complex earthquake regime with $\lambda = 1.9341$ for the simulation
    with $a/b = 0.4$, $W/L_\mathrm{nuc} = 8$, $D/L_\mathrm{nuc} = 0.1$, and
    $L/L_f = 0.5$.}
    \label{fig:S3}
\end{figure}

\begin{figure}[htbp]
    \centering
    \includegraphics[width=0.75\textwidth]{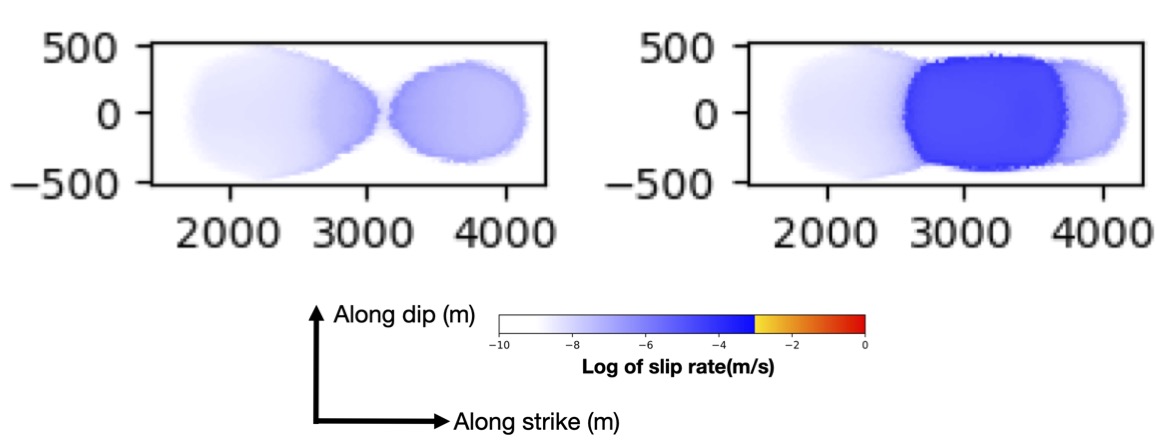}
    \caption{Slip-rate evolution on the fault plane illustrating the coalescence
    of two slow-slip events.}
    \label{fig:S4}
\end{figure}

\begin{figure}[htbp]
    \centering
    \includegraphics[width=0.85\textwidth]{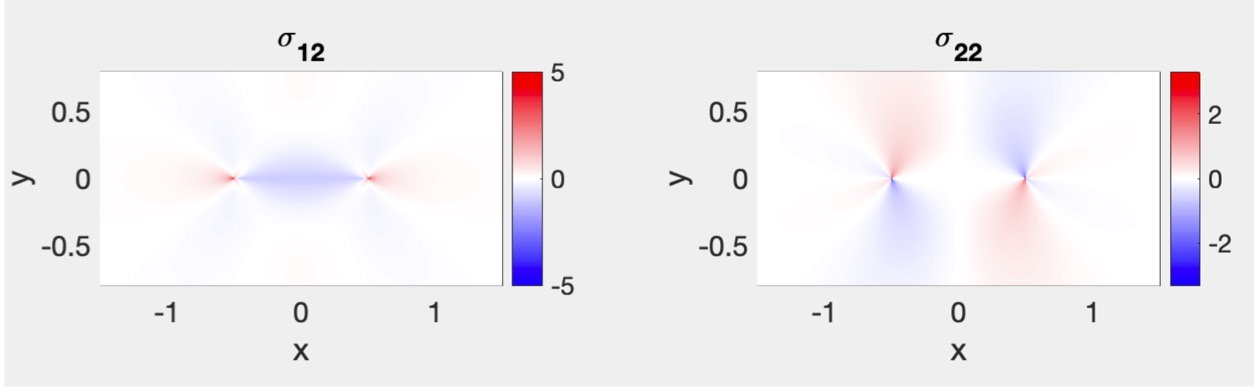}
    \caption{Shear ($\sigma_{12}$) and normal ($\sigma_{22}$) stress fields
    produced by a 2D crack subjected to a unit stress drop.}
    \label{fig:S5}
\end{figure}

\begin{figure}[htbp]
    \centering
    \includegraphics[width=0.65\textwidth]{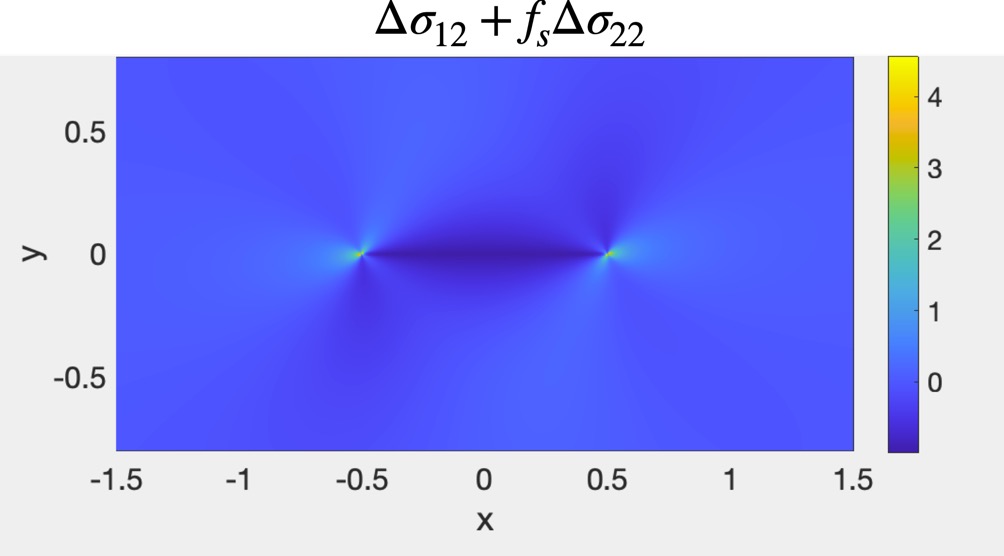}
    \caption{Coulomb stress change $\Delta\sigma_{12} + f_s \Delta \sigma_{22}$ assuming a
    friction coefficient of $f_s = 0.6$.}
    \label{fig:S6}
\end{figure}

\begin{figure}[h!]
    \centering
    \includegraphics[width=0.65\textwidth]{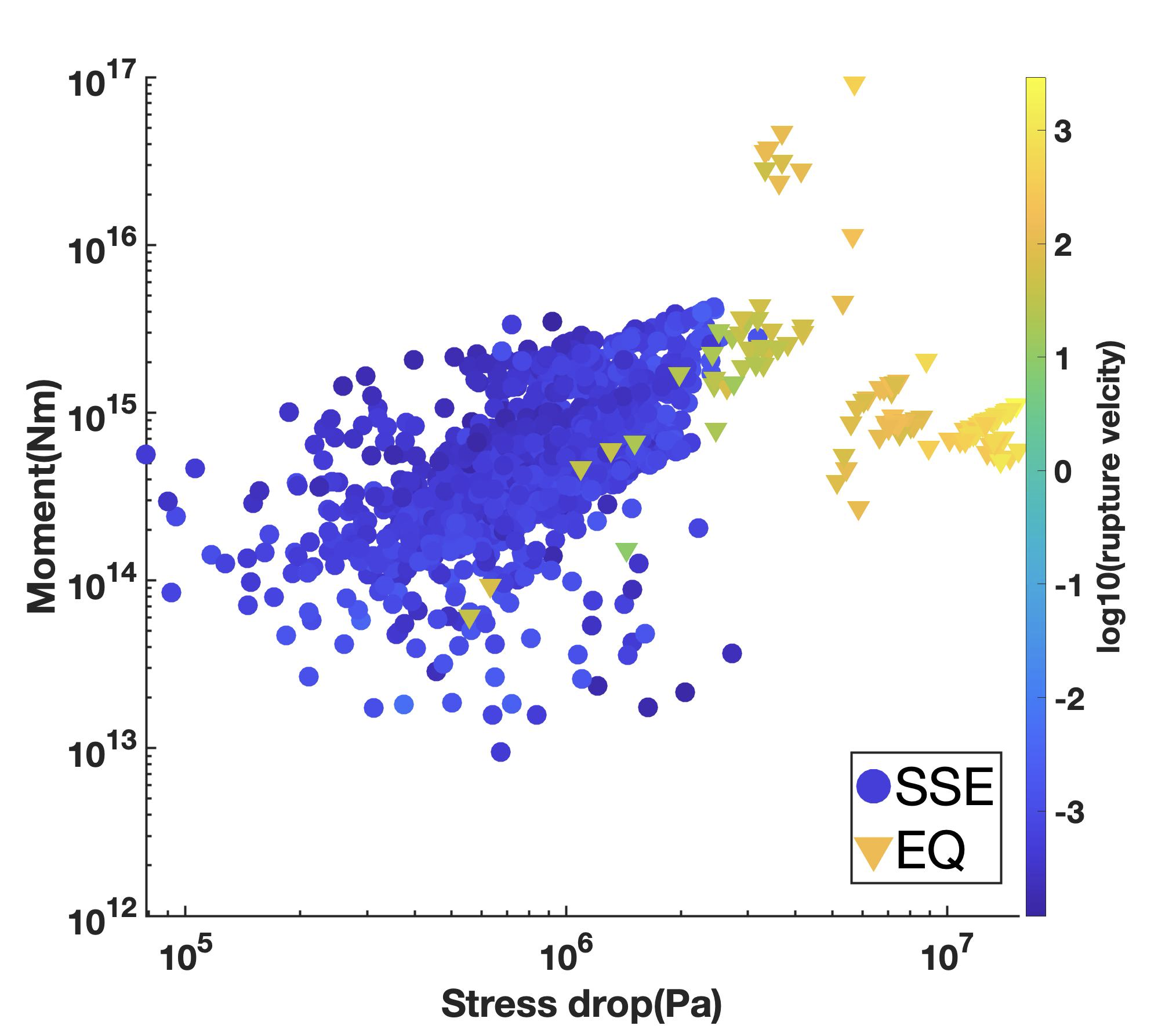}
    \caption{Stress drop and moment for slow slip events and earthquakes}
    \label{fig:S7}
\end{figure}

\clearpage
\renewcommand*{\bibfont}{\small}
\printbibliography

@misc{Viesca2026,
	author = {Viesca, Robert C. and Garagash, Dmitry I.},
	copyright = {Creative Commons Attribution Share Alike 4.0 International},
	doi = {10.48550/arXiv.2512.18729},
	publisher = {arXiv},
	title = {Linear and nonlinear stability of rate-and-state faults},
	year = {2026}}

@article{RodriguezPadilla2026,
	author = {Rodriguez Padilla, Alba Mar and Saez, Alexis and Avouac, Jean-Philippe},
	doi = {10.22541/essoar.15004063/v1},
	month = May,
	publisher = {Wiley},
	title = {Earthquake sequence synchronization and alternation driven by quasistatic elastic interactions across faults},
	year = {2026}}

@article{Mia2023,
	author = {Mia, Md Shumon and Abdelmeguid, Mohamed and Elbanna, Ahmed E.},
	doi = {10.1016/j.epsl.2023.118310},
	issn = {0012-821X},
	journal = {Earth and Planetary Science Letters},
	month = Oct,
	pages = {118310},
	publisher = {Elsevier BV},
	title = {The spectrum of fault slip in elastoplastic fault zones},
	volume = {619},
	year = {2023}}

@article{Zhai2025,
	author = {Zhai, Peng and Huang, Yihe and Liang, Chao and Ampuero, Jean-Paul},
	doi = {10.1093/gji/ggaf274},
	issn = {1365-246X},
	journal = {Geophysical Journal International},
	number = {3},
	publisher = {Oxford University Press (OUP)},
	title = {Fully dynamic seismic cycle simulations in co-evolving fault damage zones controlled by damage rheology},
	volume = {242},
	year = {2025}}

@article{Abdelmeguid2024,
	author = {Abdelmeguid, Mohamed and Mia, Md Shumon and Elbanna, Ahmed},
	doi = {10.1029/2023gl108060},
	issn = {1944-8007},
	journal = {Geophysical Research Letters},
	number = {14},
	publisher = {American Geophysical Union (AGU)},
	title = {On the Interplay Between Distributed Bulk Plasticity and Local Fault Slip in Evolving Fault Zone Complexity},
	volume = {51},
	year = {2024}}

@article{Costantino2026,
	author = {Costantino, Giuseppe and Radiguet, Mathilde and El Yousfi, Zaccaria and Socquet, Anne},
	doi = {10.1029/2025GL117446},
	journal = {Geophys. Res. Lett.},
	number = {e2025GL117446.},
	publisher = {Wiley},
	title = {A Continuum of Slow Slip Events in the Cascadia Subduction Zone Illuminated by High-Resolution Deep-Learning Denoising},
	volume = {53},
	year = {2026}}

@misc{cheng2026b,
	author = {Cheng, Jinhui and Bhat, H. S. and Almakari, M. and Lecampion, Brice and Dubernet, Pierpaolo},
	doi = {10.5281/zenodo.18168061},
	howpublished = {Zenodo},
	title = {Quantifying the Role of 3D Fault Geometry Complexities on Slow and Fast Earthquakes},
	year = {2026}}

@article{almakari2026,
	archiveprefix = {arXiv},
	author = {Almakari, Michelle and Kheirdast, N. and Villafuerte, C. D. and Thomas, M. Y. and Dubernet, Pierpaolo and Cheng, Jinhui and Gupta, Ankit and Romanet, Pierre and Chaillat, St{\'e}phanie and Bhat, Harsha S.},
	doi = {10.48550/arXiv.2509.04909},
	eprint = {2509.04909},
	journal = {under review J. Geophys. Res.},
	title = {Fault volume digital twin to reproduce the full slip spectrum, scaling and statistical laws},
	year = {2026}}

@article{Sun2025,
	author = {Sun, Yudong and Cattania, Camilla},
	doi = {10.1029/2024jb029384},
	issn = {2169-9356},
	journal = {Journal of Geophysical Research: Solid Earth},
	month = feb,
	number = {2},
	publisher = {American Geophysical Union (AGU)},
	title = {Propagation of Slow Slip Events on Rough Faults: Clustering, Back Propagation, and Re‐Rupturing},
	volume = {130},
	year = {2025}}

@article{RodriguezPiceda2025,
	author = {Rodriguez Piceda, Constanza and Mildon, Zo{\"e} K. and van den Ende, Martijn and Ampuero, Jean‐Paul and Andrews, Billy J.},
	doi = {10.1029/2024jb030382},
	issn = {2169-9356},
	journal = {Journal of Geophysical Research: Solid Earth},
	month = apr,
	number = {4},
	publisher = {American Geophysical Union (AGU)},
	title = {Normal Fault Interactions in Seismic Cycles and the Impact of Fault Network Geometry},
	url = {http://dx.doi.org/10.1029/2024JB030382},
	volume = {130},
	year = {2025}}

@article{Cheng2025,
	author = {Cheng, Jinhui and Bhat, Harsha S and Almakari, Michelle and Lecampion, Brice and Peruzzo, Carlo},
	doi = {10.1093/gji/ggaf230},
	issn = {1365-246X},
	journal = {Geophysical Journal International},
	month = jun,
	number = {2},
	publisher = {Oxford University Press (OUP)},
	title = {FASTDASH: an implementation of 3-D earthquake cycle simulation on complex fault systems using the boundary element method accelerated by H-matrices},
	url = {http://dx.doi.org/10.1093/gji/ggaf230},
	volume = {242},
	year = {2025}}

@article{Weng2025,
	author = {Weng, Huihui},
	doi = {10.1029/2024jb030663},
	issn = {2169-9356},
	journal = {Journal of Geophysical Research: Solid Earth},
	month = jun,
	number = {6},
	publisher = {American Geophysical Union (AGU)},
	title = {The Dynamics of Fast and Slow Earthquake Ruptures in Viscoelastic Materials},
	url = {http://dx.doi.org/10.1029/2024JB030663},
	volume = {130},
	year = {2025}}

@article{ElYousfi2023,
	author = {El Yousfi, Zaccaria and Radiguet, Mathilde and Rousset, Baptiste and Husker, Allen and Kazachkina, Ekaterina and Kostoglodov, Vladimir},
	doi = {10.1016/j.epsl.2023.118340},
	issn = {0012-821X},
	journal = {Earth and Planetary Science Letters},
	month = oct,
	pages = {118340},
	publisher = {Elsevier BV},
	title = {Intermittence of transient slow slip in the Mexican subduction zone},
	url = {http://dx.doi.org/10.1016/j.epsl.2023.118340},
	volume = {620},
	year = {2023}}

@article{Lecampion_BigWham_a_C_2025,
	author = {Lecampion, Brice and Fayard, Fran{\c c}ois and Gupta, Ankit and Peruzzo, Carlo and S{\'a}ez, Alexis and Richart, Nicolas and Nikolskiy, Dmitry and Ciardo, Federico},
	doi = {10.5281/zenodo.14906635},
	journal = {Zenodo},
	month = mar,
	publisher = {Zenodo},
	title = {{BigWham: a C++ library for vectorial Boundary InteGral equations With HierArchical Matrices (0.2.0)}},
	version = {0.2.0},
	year = {2025}}

@article{Shelly2009,
	author = {Shelly, David R.},
	doi = {10.1029/2009gl039589},
	issn = {1944-8007},
	journal = {Geophysical Research Letters},
	month = sep,
	number = {17},
	publisher = {American Geophysical Union (AGU)},
	title = {Possible deep fault slip preceding the 2004 Parkfield earthquake, inferred from detailed observations of tectonic tremor},
	volume = {36},
	year = {2009}}

@article{MartnezGarzn2024,
	author = {Mart{\'\i}nez-Garz{\'o}n, Patricia and Poli, Piero},
	doi = {10.1038/s43247-024-01285-y},
	issn = {2662-4435},
	journal = {Communications Earth \& Environment},
	month = mar,
	number = {1},
	publisher = {Springer Science and Business Media LLC},
	title = {Cascade and pre-slip models oversimplify the complexity of earthquake preparation in nature},
	volume = {5},
	year = {2024}}

@article{Kaneko2010,
	author = {Kaneko, Yoshihiro and Avouac, Jean-Philippe and Lapusta, Nadia},
	doi = {10.1038/ngeo843},
	issn = {1752-0908},
	journal = {Nature Geoscience},
	month = apr,
	number = {5},
	pages = {363--369},
	publisher = {Springer Science and Business Media LLC},
	title = {Towards inferring earthquake patterns from geodetic observations of interseismic coupling},
	volume = {3},
	year = {2010}}

@article{Lay2012,
	author = {Lay, Thorne and Kanamori, Hiroo and Ammon, Charles J. and Koper, Keith D. and Hutko, Alexander R. and Ye, Lingling and Yue, Han and Rushing, Teresa M.},
	doi = {10.1029/2011jb009133},
	issn = {0148-0227},
	journal = {Journal of Geophysical Research: Solid Earth},
	month = apr,
	number = {B4},
	publisher = {American Geophysical Union (AGU)},
	title = {Depth‐varying rupture properties of subduction zone megathrust faults},
	volume = {117},
	year = {2012}}

@article{Frank2019,
	author = {Frank, William B. and Brodsky, Emily E.},
	doi = {10.1126/sciadv.aaw9386},
	issn = {2375-2548},
	journal = {Science Advances},
	month = oct,
	number = {10},
	publisher = {American Association for the Advancement of Science (AAAS)},
	title = {Daily measurement of slow slip from low-frequency earthquakes is consistent with ordinary earthquake scaling},
	volume = {5},
	year = {2019}}

@article{Bletery2020,
	author = {Bletery, Quentin and Nocquet, Jean-Mathieu},
	doi = {10.1038/s41467-020-15494-4},
	issn = {2041-1723},
	journal = {Nature Communications},
	month = may,
	number = {1},
	publisher = {Springer Science and Business Media LLC},
	title = {Slip bursts during coalescence of slow slip events in Cascadia},
	volume = {11},
	year = {2020}}

@article{Hall2018,
	author = {Hall, K. and Houston, H. and Schmidt, D.},
	doi = {10.1029/2018gc007694},
	issn = {1525-2027},
	journal = {Geochemistry, Geophysics, Geosystems},
	month = aug,
	number = {8},
	pages = {2706--2718},
	publisher = {American Geophysical Union (AGU)},
	title = {Spatial Comparisons of Tremor and Slow Slip as a Constraint on Fault Strength in the Northern Cascadia Subduction Zone},
	volume = {19},
	year = {2018}}

@article{PerezSilva2021,
	author = {Perez‐Silva, Andrea and Li, Duo and Gabriel, Alice‐Agnes and Kaneko, Yoshihiro},
	doi = {10.1029/2021gl092968},
	issn = {1944-8007},
	journal = {Geophysical Research Letters},
	month = jul,
	number = {13},
	publisher = {American Geophysical Union (AGU)},
	title = {3D Modeling of Long‐Term Slow Slip Events Along the Flat‐Slab Segment in the Guerrero Seismic Gap, Mexico},
	volume = {48},
	year = {2021}}

@article{Chalumeau2024,
	author = {Chalumeau, Caroline and Agurto-Detzel, Hans and Rietbrock, Andreas and Frietsch, Michael and Oncken, Onno and Segovia, Monica and Galve, Audrey},
	doi = {10.1038/s41586-024-07245-y},
	issn = {1476-4687},
	journal = {Nature},
	month = apr,
	number = {8008},
	pages = {558--562},
	publisher = {Springer Science and Business Media LLC},
	title = {Seismological evidence for a multifault network at the subduction interface},
	volume = {628},
	year = {2024}}

@article{Kirkpatrick2021,
	author = {Kirkpatrick, James D. and Fagereng, {\AA}ke and Shelly, David R.},
	doi = {10.1038/s43017-021-00148-w},
	issn = {2662-138X},
	journal = {Nature Reviews Earth Environment},
	month = mar,
	number = {4},
	pages = {285--301},
	publisher = {Springer Science and Business Media LLC},
	title = {Geological constraints on the mechanisms of slow earthquakes},
	volume = {2},
	year = {2021}}

@article{Barnes2020,
	author = {Barnes, Philip M. and Wallace, Laura M. and Saffer, Demian M. and Bell, Rebecca E. and Underwood, Michael B. and Fagereng, Ake and Meneghini, Francesca and Savage, Heather M. and Rabinowitz, Hannah S. and Morgan, Julia K. and Kitajima, Hiroko and Kutterolf, Steffen and Hashimoto, Yoshitaka and Engelmann de Oliveira, Christie H. and Noda, Atsushi and Crundwell, Martin P. and Shepherd, Claire L. and Woodhouse, Adam D. and Harris, Robert N. and Wang, Maomao and Henrys, Stuart and Barker, Daniel H.N. and Petronotis, Katerina E. and Bourlange, Sylvain M. and Clennell, Michael B. and Cook, Ann E. and Dugan, Brandon E. and Elger, Judith and Fulton, Patrick M. and Gamboa, Davide and Greve, Annika and Han, Shuoshuo and H\"{u}pers, Andre and Ikari, Matt J. and Ito, Yoshihiro and Kim, Gil Young and Koge, Hiroaki and Lee, Hikweon and Li, Xuesen and Luo, Min and Malie, Pierre R. and Moore, Gregory F. and Mountjoy, Joshu J. and McNamara, David D. and Paganoni, Matteo and Screaton, Elizabeth J. and Shankar, Uma and Shreedharan, Srisharan and Solomon, Evan A. and Wang, Xiujuan and Wu, Hung-Yu and Pecher, Ingo A. and LeVay, Leah J.},
	doi = {10.1126/sciadv.aay3314},
	issn = {2375-2548},
	journal = {Science Advances},
	month = mar,
	number = {13},
	publisher = {American Association for the Advancement of Science (AAAS)},
	title = {Slow slip source characterized by lithological and geometric heterogeneity},
	volume = {6},
	year = {2020}}

@article{Kwiatek2024,
	author = {Kwiatek, Grzegorz and Mart{\'\i}nez‐Garz{\'o}n, Patricia and Goebel, Thomas and Bohnhoff, Marco and Ben‐Zion, Yehuda and Dresen, Georg},
	doi = {10.1029/2023jb028411},
	issn = {2169-9356},
	journal = {Journal of Geophysical Research: Solid Earth},
	month = mar,
	number = {3},
	publisher = {American Geophysical Union (AGU)},
	title = {Intermittent Criticality Multi‐Scale Processes Leading to Large Slip Events on Rough Laboratory Faults},
	volume = {129},
	year = {2024}}

@article{Ando2023,
	author = {Ando, Ryosuke and Ujiie, Kohtaro and Nishiyama, Naoki and Mori, Yasushi},
	doi = {10.1029/2022gl101388},
	issn = {1944-8007},
	journal = {Geophysical Research Letters},
	month = mar,
	number = {5},
	publisher = {American Geophysical Union (AGU)},
	title = {Depth‐Dependent Slow Earthquakes Controlled by Temperature Dependence of Brittle‐Ductile Transitional Rheology},
	volume = {50},
	year = {2023}}

@article{Bernaudin2018,
	author = {Bernaudin, M. and Gueydan, F.},
	doi = {10.1029/2018gl077586},
	issn = {1944-8007},
	journal = {Geophysical Research Letters},
	month = apr,
	number = {8},
	pages = {3471--3480},
	publisher = {American Geophysical Union (AGU)},
	title = {Episodic Tremor and Slip Explained by Fluid‐Enhanced Microfracturing and Sealing},
	volume = {45},
	year = {2018}}

@article{Wang2020,
	author = {Wang, Lifeng and Barbot, Sylvain},
	doi = {10.1126/sciadv.abb2057},
	issn = {2375-2548},
	journal = {Science Advances},
	month = sep,
	number = {36},
	publisher = {American Association for the Advancement of Science (AAAS)},
	title = {Excitation of San Andreas tremors by thermal instabilities below the seismogenic zone},
	volume = {6},
	year = {2020}}

@article{Ozawa2001,
	author = {Ozawa, Shinzaburo and Murakami, Makoto and Tada, Takashi},
	doi = {10.1029/2000jb900317},
	issn = {0148-0227},
	journal = {Journal of Geophysical Research: Solid Earth},
	month = jan,
	number = {B1},
	pages = {787--802},
	publisher = {American Geophysical Union (AGU)},
	title = {Time‐dependent inversion study of the slow thrust event in the Nankai trough subduction zone, southwestern Japan},
	volume = {106},
	year = {2001}}

@article{Nie2021,
	author = {Nie, Shiying and Barbot, Sylvain},
	doi = {10.1016/j.epsl.2021.117037},
	issn = {0012-821X},
	journal = {Earth and Planetary Science Letters},
	month = sep,
	pages = {117037},
	publisher = {Elsevier BV},
	title = {Seismogenic and tremorgenic slow slip near the stability transition of frictional sliding},
	volume = {569},
	year = {2021}}

@article{Gao2017,
	author = {Gao, Xiang and Wang, Kelin},
	doi = {10.1038/nature21389},
	issn = {1476-4687},
	journal = {Nature},
	month = mar,
	number = {7645},
	pages = {416--419},
	publisher = {Springer Science and Business Media LLC},
	title = {Rheological separation of the megathrust seismogenic zone and episodic tremor and slip},
	volume = {543},
	year = {2017}}

@article{Bhattacharya2019,
	author = {Bhattacharya, Pathikrit and Viesca, Robert C.},
	doi = {10.1126/science.aaw7354},
	issn = {1095-9203},
	journal = {Science},
	month = may,
	number = {6439},
	pages = {464--468},
	publisher = {American Association for the Advancement of Science (AAAS)},
	title = {Fluid-induced aseismic fault slip outpaces pore-fluid migration},
	volume = {364},
	year = {2019}}

@article{Wallace2018,
	author = {Wallace, Laura M. and Hreinsd{\'o}ttir, Sigr{\'u}n and Ellis, Susan and Hamling, Ian and D'Anastasio, Elisabetta and Denys, Paul},
	doi = {10.1002/2018gl077385},
	issn = {1944-8007},
	journal = {Geophysical Research Letters},
	month = may,
	number = {10},
	pages = {4710--4718},
	publisher = {Wiley},
	title = {Triggered Slow Slip and Afterslip on the Southern Hikurangi Subduction Zone Following the Kaik{\=o}ura Earthquake},
	volume = {45},
	year = {2018}}

@article{Tymofyeyeva2019,
	author = {Tymofyeyeva, Ekaterina and Fialko, Yuri and Jiang, Junle and Xu, Xiaohua and Sandwell, David and Bilham, Roger and Rockwell, Thomas K. and Blanton, Chelsea and Burkett, Faith and Gontz, Allen and Moafipoor, Shahram},
	doi = {10.1029/2018jb016765},
	issn = {2169-9356},
	journal = {Journal of Geophysical Research: Solid Earth},
	month = sep,
	number = {9},
	pages = {9956--9975},
	publisher = {American Geophysical Union (AGU)},
	title = {Slow Slip Event On the Southern San Andreas Fault Triggered by the 2017 Mw8.2 Chiapas (Mexico) Earthquake},
	volume = {124},
	year = {2019}}

@article{Uchida2019,
	author = {Uchida, Naoki},
	doi = {10.1186/s40645-019-0284-z},
	issn = {2197-4284},
	journal = {Progress in Earth and Planetary Science},
	month = may,
	number = {1},
	publisher = {Springer Science and Business Media LLC},
	title = {Detection of repeating earthquakes and their application in characterizing slow fault slip},
	volume = {6},
	year = {2019}}

@article{Michel2018,
	author = {Michel, Sylvain and Gualandi, Adriano and Avouac, Jean-Philippe},
	doi = {10.1007/s00024-018-1991-x},
	issn = {1420-9136},
	journal = {Pure and Applied Geophysics},
	month = sep,
	number = {9},
	pages = {3867--3891},
	publisher = {Springer Science and Business Media LLC},
	title = {Interseismic Coupling and Slow Slip Events on the Cascadia Megathrust},
	volume = {176},
	year = {2018}}

@article{Rousset2019,
	author = {Rousset, Baptiste and B\"{u}rgmann, Roland and Campillo, Michel},
	doi = {10.1126/sciadv.aav3274},
	issn = {2375-2548},
	journal = {Science Advances},
	month = feb,
	number = {2},
	publisher = {American Association for the Advancement of Science (AAAS)},
	title = {Slow slip events in the roots of the San Andreas fault},
	volume = {5},
	year = {2019}}

@article{Im2021a,
	author = {Im, Kyungjae and Avouac, Jean-Philippe},
	doi = {10.1016/j.geothermics.2021.102238},
	issn = {0375-6505},
	journal = {Geothermics},
	month = dec,
	pages = {102238},
	publisher = {Elsevier BV},
	title = {On the role of thermal stress and fluid pressure in triggering seismic and aseismic faulting at the Brawley Geothermal Field, California.},
	volume = {97},
	year = {2021}}

@article{scheel2022,
	author = {Scheel, Johannes and Wallenta, Daniel and Ricoeur, Andreas},
	doi = {10.1007/s10659-022-09883-7},
	journal = {J. Elasticity},
	number = {1-2},
	pages = {291--308},
	title = {A Critical Review on the Complex Potentials in Linear Elastic Fracture Mechanics},
	volume = {147},
	year = 2022}

@article{burgmann2018,
	author = {B{\"u}rgmann, Roland},
	doi = {10.1016/j.epsl.2018.04.062},
	journal = {Earth Planet. Sc. Lett.},
	pages = {112--134},
	title = {The geophysics, geology and mechanics of slow fault slip},
	volume = {495},
	year = {2018}}

@article{avouac2015,
	author = {Avouac, Jean-Philippe},
	doi = {10.1146/annurev-earth-060614-105302},
	journal = {Ann. Rev. Earth Planet. Sci.},
	pages = {233--271},
	title = {From geodetic imaging of seismic and aseismic fault slip to dynamic modeling of the seismic cycle},
	volume = {43},
	year = {2015}}

@article{cruz2018,
	author = {Cruz-Atienza, V. M. and Villafuerte, C. D. and Bhat, H. S.},
	doi = {10.1038/s41467-018-05150-3},
	journal = {Nat. Commun.},
	number = {1},
	pages = {2900},
	pubnumber = {29},
	title = {Rapid tremor migration and pore-pressure waves in subduction zones},
	topics = {tremor, permeability},
	volume = {9},
	year = {2018}}

@article{luo2017b,
	author = {Luo, Yingdi and Ampuero, Jean-Paul},
	doi = {10.1016/j.tecto.2017.11.006},
	journal = {Tectonophysics},
	title = {Stability of faults with heterogeneous friction properties and effective normal stress},
	year = {2017}}

@article{lapusta2009a,
	author = {Lapusta, Nadia and Liu, Yi},
	doi = {10.1029/2008JB005934},
	journal = {J. Geophys. Res.},
	number = {B9},
	publisher = {Wiley Online Library},
	title = {Three-dimensional boundary integral modeling of spontaneous earthquake sequences and aseismic slip},
	volume = {114},
	year = {2009}}

@article{wallace2013,
	author = {Wallace, Laura M and Eberhart-Phillips, Donna},
	doi = {10.1002/2013GL057682},
	journal = {Geophys. Res. Lett.},
	number = {20},
	pages = {5393--5398},
	publisher = {Wiley Online Library},
	title = {Newly observed, deep slow slip events at the central Hikurangi margin, New Zealand: Implications for downdip variability of slow slip and tremor, and relationship to seismic structure},
	volume = {40},
	year = {2013}}

@article{ito2007,
	author = {Ito, Yoshihiro and Obara, Kazushige and Shiomi, Katsuhiko and Sekine, Shutaro and Hirose, Hitoshi},
	doi = {10.1126/science.1134454},
	journal = {Science},
	number = {5811},
	pages = {503--506},
	publisher = {American Association for the Advancement of Science},
	title = {Slow earthquakes coincident with episodic tremors and slow slip events},
	volume = {315},
	year = {2007}}

@article{hirose2006,
	author = {Hirose, Hitoshi and Obara, Kazushige},
	doi = {10.1029/2006GL026579},
	journal = {Geophys. Res. Lett.},
	number = {17},
	publisher = {Wiley Online Library},
	title = {Short-term slow slip and correlated tremor episodes in the Tokai region, central Japan},
	volume = {33},
	year = {2006}}

@article{mitsui2006,
	author = {Mitsui, Noa and Hirahara, Kazuro},
	doi = {10.1016/j.epsl.2006.03.001},
	journal = {Earth Planet. Sc. Lett.},
	number = {1},
	pages = {344--358},
	publisher = {Elsevier},
	title = {Slow slip events controlled by the slab dip and its lateral change along a trench},
	volume = {245},
	year = {2006}}

@article{liliu2016,
	author = {Li, Duo and Liu, Yajing},
	doi = {10.1002/2016JB012857},
	journal = {J. Geophys. Res.},
	number = {9},
	pages = {6828--6845},
	publisher = {Wiley Online Library},
	title = {Spatiotemporal evolution of slow slip events in a nonplanar fault model for northern Cascadia subduction zone},
	volume = {121},
	year = {2016}}

@article{romanet2018,
	author = {Romanet, Pierre and Bhat, H. S. and Jolivet, Romain and Madariaga, Ra{\`u}l},
	doi = {10.1029/2018GL077579},
	journal = {Geophys. Res. Lett.},
	pubnumber = {27},
	title = {Fast and slow earthquakes emerge due to fault geometrical complexity},
	topics = {numerical models, BIEM, H- Matrices, Slow Slip, earthquake cycles},
	year = {2018}}

@article{nakata2011,
	author = {Nakata, Ryoko and Ando, Ryosuke and Hori, Takane and Ide, Satoshi},
	doi = {10.1029/2010JB008188},
	journal = {J. Geophys. Res.},
	pages = {B08308},
	publisher = {Wiley Online Library},
	title = {Generation mechanism of slow earthquakes: Numerical analysis based on a dynamic model with brittle-ductile mixed fault heterogeneity},
	volume = {116},
	year = {2011}}

@article{bouchon2011,
	author = {Bouchon, M and Karabulut, H and Aktar, M and Ozalaybey, S and Schmittbuhl, J and Bouin, M P},
	doi = {10.1126/science.1197341},
	journal = {Science},
	number = {6019},
	pages = {877--880},
	title = {{Extended Nucleation of the 1999 Mw 7.6 Izmit Earthquake}},
	volume = {331},
	year = {2011}}

@article{obara2016,
	author = {Obara, Kazushige and Kato, Aitaro},
	doi = {10.1126/science.aaf1512},
	journal = {Science},
	number = {6296},
	pages = {253--257},
	publisher = {American Association for the Advancement of Science},
	title = {Connecting slow earthquakes to huge earthquakes},
	volume = {353},
	year = {2016}}

@article{liu2010,
	author = {Liu, Yajing and Rubin, Allan M},
	doi = {10.1029/2010JB007522},
	journal = {J. Geophys. Res.},
	number = {B10},
	publisher = {Wiley Online Library},
	title = {Role of fault gouge dilatancy on aseismic deformation transients},
	volume = {115},
	year = {2010}}

@article{obara2004,
	author = {Obara, Kazushige and Hirose, Hitoshi and Yamamizu, Fumio and Kasahara, Keiji},
	doi = {10.1029/2004GL020848},
	journal = {Geophys. Res. Lett.},
	number = {23},
	publisher = {Wiley Online Library},
	title = {Episodic slow slip events accompanied by non-volcanic tremors in southwest Japan subduction zone},
	volume = {31},
	year = {2004}}

@article{jolivet2013,
	author = {Jolivet, Romain and Lasserre, C{\'e}cile and Doin, M-P and Peltzer, G and Avouac, J-P and Sun, Jianbao and Dailu, R},
	doi = {10.1016/j.epsl.2013.07.020},
	journal = {Earth Planet. Sc. Lett.},
	pages = {23--33},
	publisher = {Elsevier},
	title = {Spatio-temporal evolution of aseismic slip along the Haiyuan fault, China: Implications for fault frictional properties},
	volume = {377},
	year = {2013}}

@article{dragert2001,
	author = {Dragert, Herb and Wang, Kelin and James, Thomas S},
	journal = {Science},
	number = {5521},
	pages = {1525--1528},
	publisher = {American Association for the Advancement of Science},
	title = {A silent slip event on the deeper Cascadia subduction interface},
	volume = {292},
	year = {2001}}

@article{hirose1999,
	author = {Hirose, Hitoshi and Hirahara, Kazuro and Kimata, Fumiaki and Fujii, Naoyuki and Miyazaki, Shin'ichi},
	journal = {Geophys. Res. Lett.},
	number = {21},
	pages = {3237--3240},
	publisher = {Wiley Online Library},
	title = {A slow thrust slip event following the two 1996 Hyuganada earthquakes beneath the Bungo Channel, southwest Japan},
	volume = {26},
	year = {1999}}

@article{radiguet2016,
	author = {Radiguet, M. and Perfettini, H. and Cotte, N. and Gualandi, A. and Valette, B. and Kostoglodov, V. and Lhomme, T. and Walpersdorf, A. and Cabral Cano, E. and Campillo, M.},
	doi = {10.1038/NGEO2817},
	journal = {Nature Geoscience},
	pages = {829--833},
	title = {Triggering of the 2014 Mw7.3 Papanoa earthquake by a slow slip event in Guerrero, Mexico},
	volume = {9},
	year = {2016}}

@article{viesca2016b,
	author = {Viesca, R. C.},
	doi = {10.1098/rspa.2016.0254},
	journal = {Proc. R. Soc. A},
	title = {Self-similar slip instability on interfaces with rate- and state-dependent friction},
	year = {2016}}

@article{kato2003,
	author = {Kato, N},
	journal = {Bull. Earthq. Res. Inst.},
	pages = {151--166},
	title = {Repeating slip events at a circular asperity: numerical simulation with a rate- and state-dependent friction law},
	volume = {78},
	year = {2003}}

@article{chen2009,
	author = {Chen, Ting and Lapusta, Nadia},
	doi = {10.1029/2008JB005749},
	journal = {J. Geophys. Res.},
	number = {B1},
	publisher = {Wiley Online Library},
	title = {Scaling of small repeating earthquakes explained by interaction of seismic and aseismic slip in a rate and state fault model},
	volume = {114},
	year = {2009}}

@article{skarbek2012,
	author = {Skarbek, RM and Rempel, AW and Schmidt, DA},
	doi = {10.1029/2012GL053762},
	journal = {Geophys. Res. Lett.},
	number = {21},
	publisher = {Wiley Online Library},
	title = {Geologic heterogeneity can produce aseismic slip transients},
	volume = {39},
	year = {2012}}

@article{rubin2008,
	author = {Rubin, Allan M},
	doi = {10.1029/2008JB005642},
	journal = {J. Geophys. Res.},
	pages = {B11414},
	publisher = {Wiley Online Library},
	title = {Episodic slow slip events and rate-and-state friction},
	volume = {113},
	year = {2008}}

@article{veedu2016,
	author = {Veedu, Deepa Mele and Barbot, Sylvain},
	doi = {10.1038/nature17190},
	journal = {Nature},
	number = {7599},
	pages = {361--365},
	publisher = {Nature Publishing Group},
	title = {The Parkfield tremors reveal slow and fast ruptures on the same asperity},
	volume = {532},
	year = {2016}}

@article{wallace2016,
	author = {Wallace, Laura M and Webb, Spahr C and Ito, Yoshihiro and Mochizuki, Kimihiro and Hino, Ryota and Henrys, Stuart and Schwartz, Susan Y and Sheehan, Anne F},
	journal = {Science},
	number = {6286},
	pages = {701--704},
	publisher = {American Association for the Advancement of Science},
	title = {Slow slip near the trench at the Hikurangi subduction zone, New Zealand},
	volume = {352},
	year = {2016}}

@article{dublanchet2013,
	author = {Dublanchet, Pierre and Bernard, Pascal and Favreau, Pascal},
	doi = {10.1002/jgrb.50187},
	journal = {J. Geophys. Res.},
	number = {5},
	pages = {2225--2245},
	publisher = {Wiley Online Library},
	title = {Interactions and triggering in a 3-D rate-and-state asperity model},
	volume = {118},
	year = {2013}}

@article{liu2007,
	author = {Liu, Yajing and Rice, James R},
	doi = {10.1029/2007JB004930},
	journal = {J. Geophys. Res.},
	pages = {B09404},
	publisher = {Wiley Online Library},
	title = {Spontaneous and triggered aseismic deformation transients in a subduction fault model},
	volume = {112},
	year = {2007}}

@article{segall2010a,
	author = {Segall, P. and Rubin, A. M. and Bradley, A. M. and Rice, J. R.},
	doi = {10.1029/2010JB007449,},
	journal = {J. Geophys. Res.},
	pages = {B12305},
	publisher = {American Geophysical Union},
	title = {Dilatant strengthening as a mechanism for slow slip events},
	volume = {115},
	year = {2010}}

@article{shibazaki2007,
	author = {Shibazaki, B. and Shimamoto, T.},
	doi = {10.1111/j.1365-246X.2007.03434.x},
	journal = {Geophys. J. Int.},
	number = {1},
	pages = {191--205},
	publisher = {Wiley Online Library},
	title = {Modelling of short-interval silent slip events in deeper subduction interfaces considering the frictional properties at the unstable--stable transition regime},
	volume = {171},
	year = {2007}}

@article{kato2012,
	author = {Kato, A. and Obara, K. and Igarashi, T. and Tsuruoka, H. and Nakagawa, S. and Hirata, N.},
	doi = {10.1126/science.1215141},
	journal = {Science},
	number = {6069},
	pages = {705--708},
	title = {{Propagation of Slow Slip Leading Up to the 2011 Mw 9.0 Tohoku-Oki Earthquake}},
	volume = {335},
	year = {2012}}

@article{shelly2007a,
	author = {Shelly, D. R. and Beroza, G. C. and Ide, S.},
	doi = {10.1038/nature05666},
	issn = {0028-0836},
	journal = {Nature},
	number = {7133},
	pages = {305--307},
	publisher = {Nature Publishing Group},
	title = {{Non-volcanic tremor and low-frequency earthquake swarms}},
	volume = {446},
	year = {2007}}

@article{rogers2003,
	author = {Rogers, G. and Dragert, H.},
	journal = {Science},
	number = {5627},
	pages = {1942--1943},
	publisher = {AAAS},
	title = {{Episodic tremor and slip on the Cascadia subduction zone: The chatter of silent slip}},
	volume = {300},
	year = {2003}}

@article{liu2005,
	author = {Liu, Y. and Rice, J. R.},
	doi = {10.1029/2004JB003424},
	journal = {J. Geophys. Res.},
	pages = {B08307},
	title = {Aseismic slip transients emerge spontaneously in 3D rate and state modeling of subduction earthquake sequences},
	volume = {110},
	year = {2005}}

@article{rice1993,
	author = {Rice, J. R.},
	doi = {10.1029/93JB00191},
	journal = {J. Geophys. Res.},
	number = {B6},
	pages = {9885--9907},
	title = {Spatio-temporal complexity of slip on a fault},
	volume = {98},
	year = {1993}}

@article{rubin2005,
	author = {Rubin, AM and Ampuero, J-P},
	doi = {10.1029/2005JB003686},
	journal = {J. Geophys. Res.},
	pages = {B11312},
	publisher = {Wiley Online Library},
	title = {Earthquake nucleation on (aging) rate and state faults},
	volume = {110},
	year = {2005}}

@article{Cattania2021,
	author = {Cattania, Camilla and Segall, Paul},
	doi = {10.1029/2020jb020430},
	journal = {Journal of Geophysical Research: Solid Earth},
	month = apr,
	number = {4},
	publisher = {American Geophysical Union ({AGU})},
	title = {Precursory Slow Slip and Foreshocks on Rough Faults},
	volume = {126},
	year = {2021}}
\end{document}